# On Achievable Rates and Complexity of LDPC Codes for Parallel Channels with Application to Puncturing

Igal Sason    Gil Wiechman

Technion – Israel Institute of Technology
Haifa 32000, Israel
{sason@ee, igillw@tx}.technion.ac.il

July 26, 2018

**Abstract**

This paper considers the achievable rates and decoding complexity of low-density parity-check (LDPC) codes over statistically independent parallel channels. The paper starts with the derivation of bounds on the conditional entropy of the transmitted codeword given the received sequence at the output of the parallel channels; the component channels are considered to be memoryless, binary-input, and output-symmetric (MBIOS). These results serve for the derivation of an upper bound on the achievable rates of ensembles of LDPC codes under optimal maximum-likelihood (ML) decoding when their transmission takes place over parallel MBIOS channels. The paper relies on the latter bound for obtaining upper bounds on the achievable rates of ensembles of randomly and intentionally punctured LDPC codes over MBIOS channels. For ensembles of punctured LDPC codes, the calculation of bounds on their thresholds under ML decoding and their exact thresholds under iterative decoding (based on density evolution analysis) is of interest in the sense that it enables to assess the degradation in the asymptotic performance which is attributed to the sub-optimality of iterative decoding (as compared to ML decoding), and also to assess the inherent loss in the asymptotic performance which is attributed to the structure of these ensembles, even if ML decoding could be applied to decode LDPC codes. The paper also provides a lower bound on the decoding complexity (per iteration) of ensembles of LDPC codes under message-passing iterative decoding over parallel MBIOS channels; the bound is given in terms of the gap between the rate of these codes for which reliable communication is achievable and the channel capacity. Similarly to the case of a single MBIOS channel, this lower bound grows like the log of the inverse of the gap to capacity. The latter bound is used for the derivation of lower bounds on the decoding complexity of punctured LDPC codes over MBIOS channels; looser versions of these bounds suggest a simplified re-derivation of previously reported bounds on the decoding complexity of randomly punctured LDPC codes. The paper presents a diagram which shows interconnections between the theorems introduced in this paper and some other previously reported results. The setting which serves for the derivation of the bounds on the achievable rates and decoding complexity is general, and the bounds can be applied to punctured LDPC codes, non-uniformly error protected LDPC codes, and LDPC-coded modulation where all these scenarios can be treated as different forms of communication over parallel channels.

# 1 Introduction

This paper focuses primarily on the information-theoretic limitations of low-density parity-check (LDPC) codes whose transmission takes place over a set of parallel channels. By parallel channels we mean a communication system or model where the distortion introduced by each channel is independent of the signal and the distortion in all of the other channels. The first main result presented in this paper is the derivation of an upper bound on the achievable rates of ensembles of LDPC codes under optimal maximum-likelihood (ML) decoding when we assume that the codes are communicated over parallel memoryless binary-input output-symmetric (MBIOS) channels. This result forms a non-trivial generalization of the “un-quantized” bound derived by the authors for a single MBIOS channel [17], where the latter bound improves the tightness of the bound in [1] which was based on a two-level quantization approach. The second main result introduced in this paper is a derivation of a lower bound on the decoding complexity of ensembles of LDPC codes for parallel MBIOS channels; this bound assumes an iterative message-passing decoder, and the decoding complexity is normalized per iteration. In this respect, it continues the study in [3] which was focused on upper bounds on the parity-check density of ensembles of LDPC codes for parallel MBIOS channels. The two main issues addressed in this paper for parallel channels are applied to punctured LDPC codes.

Puncturing serves to increase the rate of the original code by reducing the length of the codeword. The performance of punctured LDPC codes under ML decoding was studied in [2] via analyzing the asymptotic growth rate of their average weight distributions and using upper bounds on the decoding error probability under ML decoding. Based on this analysis, it was proved that for any MBIOS channel, capacity-achieving codes of any rate can be constructed by puncturing the code bits of ensembles of LDPC codes with small enough design rate. In [4], the performance of punctured LDPC codes was studied under message-passing iterative decoding for the AWGN channel. Ha and McLaughlin studied in [4] two methods for puncturing LDPC codes where the first method assumes random puncturing of the code bits at a fixed rate, and the second method assumes possibly different puncturing rates for each subset of code bits which corresponds to variable nodes of a fixed degree. In the second approach which is called ’intentional puncturing’, the degree distributions of the puncturing patterns were optimized in [4, 5] were it was aimed to minimize the threshold under iterative decoding for a given design rate via the Gaussian approximation; exact values of these optimized puncturing patterns were also calculated by the density evolution analysis and show good agreement with the Gaussian approximation. The results in [4, 5] exemplify the usefulness of punctured LDPC codes for a relatively wide range of rates, and therefore, they are suitable for rate-compatible coding. Some fundamental properties of punctured LDPC codes were studied in [12], and it was specifically claimed that for an arbitrary MBIOS channel and rates $R_1$ and $R_2$ satisfying $0 < R_1 < R_2 < 1$, there exists an ensemble of LDPC codes which can be punctured from rate $R_1$ to rate $R_2$ resulting in good codes for all rates $R_1 \leq R \leq R_2$. This interesting property implies that rates arbitrarily close to one are achievable via puncturing of a suitable ensemble of LDPC codes whose rate (before puncturing) is sufficiently low. These results show the high potential of puncturing in designing codes which operate closely to the Shannon capacity limit and used for rate-compatible coding for various memoryless binary-input output-symmetric (MBIOS) channels.

The transmission of punctured codes over an arbitrary channel can be treated as a particular case of communication of the original code over a set of parallel channels (where the puncturing rate of each set of code bits cascaded with the communication channel of the punctured code serve to define the set of parallel channels for the transmission of the original code before puncturing). We therefore apply the bounds on the achievable rates and decoding complexity of LDPC codes over statistically independent parallel channels to the case of transmission of ensembles of punctured LDPC codes over a single MBIOS channel. We state puncturing theorems related to achievable rates and decoding complexity of punctured LDPC codes. For ensembles of punctured LDPC codes, the calculation of bounds on their thresholds under ML decoding and their exact thresholds under iterative decoding (based on density evolution analysis) is of interest in the sense that it enables to assess the degradation in the asymptotic performance which is attributed to the sub-optimality of iterative decoding (as compared to ML decoding), and also to assess the inherent loss in the asymptotic performance which is attributed to the structure of these ensembles, even if ML decoding could be applied to decode LDPC codes.

The paper is organized as follows: Section 2 presents preliminary material, Section 3 derives bounds on the conditional entropy of the transmitted codeword given the received sequence at the output of the parallel channels where the component channels are considered to be MBIOS. Section 4 relies on the previous bounds and derives an upper bound on the achievable rates of LDPC codes under ML decoding for parallel channels. Section 5 relies on the previous section for the derivation of upper bounds on the achievable rates of ensembles of randomly and intentionally punctured LDPC codes whose transmission takes place over MBIOS channels, and numerical results are exemplified for various ensembles. Section 6 provides a lower bound on the decoding complexity of ensembles of LDPC codes under message-passing iterative decoding for parallel MBIOS channels. The latter result is used in this section for the derivation of lower bounds on the decoding complexity of randomly and intentionally punctured LDPC codes for MBIOS channels; looser versions of these bounds suggest a simplified re-derivation of previously reported bounds on the decoding complexity of randomly punctured LDPC codes (as shown in an appendix). Finally, Section 7 summarizes our discussion, and presents a diagram which shows interconnections between the theorems introduced in this paper and some other previously reported results from [1, 10, 11, 13, 15, 17].

## 2 Preliminaries

Assume that the communication takes place over $J$ statistically independent parallel channels where each of the individual channels is an MBIOS channel whose probability density function is given by $p(\cdot|\cdot\,;j)$ $(j = 1, 2, \ldots, J)$. If we let $\mathcal{I}(j)$ be the set of indices of the symbols in an $n$-length codeword which are transmitted over the $j^{\text{th}}$ channel, then

$$p_n\left(\underline{y}|\underline{x}\right) = \prod_{j=1}^{J} \prod_{i \in \mathcal{I}(j)} p(y_i|x_i; j)\ . \tag{1}$$

In this paper, we consider the transmission of binary linear block codes over a set of $J$ parallel independent MBIOS channels. The results apply both to individual codes and ensembles, and are applied to ensembles of LDPC codes. In the following, we briefly present standard notation used for the characterization of ensembles of LDPC codes, and also present some reported results on LDPC codes which are related to this paper.

In general, an LDPC code is a linear block code which is represented by a sparse parity-check matrix $H$. This matrix can be expressed in an equivalent form by a bipartite graph $\mathcal{G}$ whose variable nodes (appearing on the left of $\mathcal{G}$) represent the the code bits, and whose parity-check nodes (appearing on the right of $\mathcal{G}$) represent the linear constraints defined by $H$. In such a bipartite graph, an edge connects a variable node with a parity-check node if and only if the variable is involved in the corresponding parity-check equation; the degree of a node is defined as the number of edges which are attached to it.

Following standard notation, let $\lambda_i$ and $\rho_i$ denote the fraction of *edges* attached to variable and parity-check nodes of degree $i$, respectively. In a similar manner, let $\Lambda_i$ and $\Gamma_i$ denote the fraction of variable and parity-check nodes of degree $i$, respectively. The LDPC ensemble is characterized by a triplet $(n, \lambda, \rho)$ where $n$ designates the block length of the codes, and the polynomials

$$\lambda(x) \triangleq \sum_{i=1}^{\infty} \lambda_i x^{i-1}, \qquad \rho(x) \triangleq \sum_{i=1}^{\infty} \rho_i x^{i-1} \tag{2}$$

represent, respectively, the left and right degree distributions (d.d.) from the *edge* perspective. Equivalently, this ensemble can be also characterized by the triplet $(n, \Lambda, \Gamma)$ where the polynomials

$$\Lambda(x) \triangleq \sum_{i=1}^{\infty} \Lambda_i x^{i}, \qquad \Gamma(x) \triangleq \sum_{i=1}^{\infty} \Gamma_i x^{i} \tag{3}$$

represent, respectively, the left and right d.d. from the *node* perspective. We denote by LDPC$(n, \lambda, \rho)$ (or LDPC$(n, \Lambda, \Gamma)$) the ensemble of codes whose bipartite graphs are constructed according to the corresponding pairs of degree distributions. One can switch between degree distributions w.r.t. to the nodes and edges of a bipartite graph, using the following equations [14]:

$$\Lambda(x) = \frac{\int_0^x \lambda(u)du}{\int_0^1 \lambda(u)du}, \qquad \Gamma(x) = \frac{\int_0^x \rho(u)du}{\int_0^1 \rho(u)du} \tag{4}$$

$$\lambda(x) = \frac{\Lambda'(x)}{\Lambda'(1)}, \qquad \rho(x) = \frac{\Gamma'(x)}{\Gamma'(1)}. \tag{5}$$

An important characteristic of an ensemble of LPDC codes is its *design rate*. For an LDPC ensemble whose codes are represented by parity-check matrices of dimension $c \times n$, the design rate is defined to be $R_{\text{d}} \triangleq 1 - \frac{c}{n}$. This serves as a lower bound on the actual rate of any code from this ensemble, and is equal to the actual rate if the parity-check matrix of a code is *full rank* (i.e., the linear constraints which define this code are linearly independent). For an ensemble of LDPC codes, the design rate can be calculated in terms of the degree distributions (either w.r.t. the edges or nodes of a graph) by two equivalent expressions:

$$R_{\text{d}} = 1 - \frac{\int_0^1 \rho(x)dx}{\int_0^1 \lambda(x)dx} = 1 - \frac{\Lambda'(1)}{\Gamma'(1)}. \tag{6}$$

A sufficient condition for the asymptotic convergence of the rate of codes from an LDPC ensemble to its design rate was recently stated in [8, lemma 7].

**Lemma 2.1 ([8], Lemma 7).** Let $\mathcal{C}$ be a code which is chosen uniformly at random from the ensemble LDPC$(n, \Lambda, \Gamma)$, let $R$ be the rate of $\mathcal{C}$, and $R_{\text{d}}$ be the design rate of this ensemble. Consider the function

$$\begin{aligned}\Psi(u) \;\triangleq\; & -\Lambda'(1) \log_2 \left[\frac{1+uv}{(1+u)(1+v)}\right] \\ & + \sum_{i=1}^{\infty} \Lambda_i \log_2 \left[\frac{1+u^i}{2(1+u)^i}\right] + \frac{\Lambda'(1)}{\Gamma'(1)} \sum_{i=1}^{\infty} \Gamma_i \log_2 \left[1 + \left(\frac{1-v}{1+v}\right)^i\right]\end{aligned}$$

where

$$v \triangleq \left( \sum_{i=1}^{\infty} \frac{\lambda_i}{1+u^i} \right)^{-1} \left( \sum_{i=1}^{\infty} \frac{\lambda_i u^{i-1}}{1+u^i} \right).$$

Assume that the function $\Psi$ achieves its global maximum in the range $u \in [0, \infty)$ at $u = 1$. Then, there exists a constant $B > 0$ such that, for any $\xi > 0$ and $n > n_0(\xi, \Lambda, \Gamma)$,

$$\Pr\{|R - R_{\mathrm{d}}| > \xi\} \leq e^{-Bn\xi} .$$

Moreover, there exists $C > 0$ such that, for $n > n_0(\xi, \Lambda, \Gamma)$

$$\mathbb{E}\{|R - R_{\mathrm{d}}|\} \leq \frac{C \log n}{n} .$$

In Section 5.4, we rely on this lemma in order to verify that the design rates of various ensembles of LDPC codes are equal in probability 1 to the asymptotic rates of codes from these ensembles.

# 3 Bounds on the Conditional Entropy for Parallel Channels

This section serves as a preparatory step towards the derivation of upper bounds on the achievable rates of ML decoded binary linear block codes whose transmission takes place over statistically independent parallel MBIOS channels. To this end, we present in this section upper and lower bounds on the conditional entropy of the transmitted codeword given the received sequence at the output of these channels.

## 3.1 Lower Bound on the Conditional Entropy

We begin by deriving an information-theoretic lower bound on the conditional entropy of the transmitted codeword given the received sequence, when the transmission takes place over a set of $J$ independent parallel MBIOS channels.

**Proposition 3.1.** Let $\mathcal{C}$ be a binary linear block code of length $n$, and assume that its transmission takes place over a set of $J$ statistically independent parallel MBIOS channels. Let $C_j$ denote the capacity of the $j^{\text{th}}$ channel (in bits per channel use), and $a(\cdot; j)$ designate the conditional *pdf* of the log-likelihood ratio (LLR) at the output of the $j^{\text{th}}$ channel given its input is 1 (where $j \in \{1, \ldots, J\}$ and the channel input can be either $\pm 1$). Let $\mathbf{x} = (x_1, \ldots x_n)$ and $\mathbf{y} = (y_1, \ldots, y_n)$ designate the transmitted codeword and received sequence, respectively, $\mathcal{I}(j)$ be the set of indices of the code bits transmitted over the $j^{\text{th}}$ channel, $n^{[j]} \triangleq |\mathcal{I}(j)|$ be the size of this set, and $p_j \triangleq \frac{n^{[j]}}{n}$ be the fraction of bits transmitted over the $j^{\text{th}}$ channel. For an arbitrary $c \times n$ parity-check matrix $H$ of the code $\mathcal{C}$, let $\beta_{j,m}$ designate the number of indices in $\mathcal{I}(j)$ referring to bits which are involved in the $m^{\text{th}}$ parity-check equation of $H$ (where $m \in \{1, \ldots, c\}$), and let $R_{\mathrm{d}} = 1 - \frac{c}{n}$ be the design rate of $\mathcal{C}$. Then, the conditional entropy of the transmitted codeword given the received sequence satisfies

$$\frac{H(\mathbf{X}|\mathbf{Y})}{n} \geq 1 - \sum_{j=1}^{J} p_j C_j - (1-R_{\mathrm{d}}) \left( 1 - \frac{1}{2n(1-R_{\mathrm{d}}) \ln 2} \sum_{p=1}^{\infty} \left\{ \frac{1}{p(2p-1)} \sum_{m=1}^{n(1-R_{\mathrm{d}})} \prod_{j=1}^{J} (g_{j,p})^{\beta_{j,m}} \right\} \right) \quad (7)$$

where

$$g_{j,p} \triangleq \int_0^{\infty} a(l;j)\ (1+e^{-l}) \tanh^{2p}\left(\frac{l}{2}\right) dl, \qquad j \in \{1, \ldots, J\}, \quad p \in \mathbb{N}. \quad (8)$$

*Proof.* The proof relies on concepts which are presented in [1, 17], and generalizes them to the case of parallel channels. Let us assume that $\mathbf{x} \in \mathcal{C}$ is the transmitted codeword, and that the '0' and '1' symbols are mapped to $+1$ and $-1$, respectively. The input alphabet of each component channel is $\{+1, -1\}$, and if a symbol is transmitted over the $j^{\text{th}}$ MBIOS channel and $y$ is its corresponding output, then the LLR gets the form

$$\text{LLR}(y; j) = \ln\left(\frac{p(y|X = 1; j)}{p(y|X = -1; j)}\right), \qquad j \in \{1, \ldots, J\}, \quad y \in \mathcal{Y}$$

where $\mathcal{Y}$ denotes the output alphabet of each component channel,[1] and $p(\cdot|\cdot; j)$ is the conditional *pdf* of the $j^{\text{th}}$ channel. For each one of these $J$ component channels, we move from the original mapping of $X \to Y$ (where according to (1), each symbol is transmitted over only one of these $J$ channels) to an equivalent representation of the channel $X \to \widetilde{Y}$, so that $H(X|\widetilde{Y}) = H(X|Y)$. The basic idea for showing the equivalence between the original channel and the one which will be introduced shortly is based on the principle that the LLR forms a sufficient statistics of an MBIOS channel, and the LLR is an anti-symmetric function of the channel output (i.e., $\text{LLR}(-y; j) = -\text{LLR}(y; j)$ for $j = 1, \ldots, J$). This yields that the absolute value of the LLR doesn't change when the channel output is flipped, but the sign of the LLR alternates.

In the following, we characterize an equivalent channel to the considered set of $J$ parallel channels. For each index $i \in \mathcal{I}(j)$, let us choose independently a value $l_i$ according to the conditional *pdf* $a(\cdot; j)$, and for $i \in \{1, \ldots, n\}$, let

$$\omega_i \triangleq |l_i|, \quad \theta_i \triangleq \begin{cases} +1 & \text{if} \quad l_i > 0 \\ -1 & \text{if} \quad l_i < 0 \\ \pm 1 \text{ w.p. } \frac{1}{2} & \text{if} \quad l_i = 0 \end{cases}.$$

The output of the equivalent channel is defined as $\widetilde{\mathbf{y}} = (\widetilde{y}_1, \ldots, \widetilde{y}_n)$ where $\widetilde{y}_i = (\phi_i, \omega_i)$ and $\phi_i = \theta_i x_i$. This defines the mapping

$$X \to \widetilde{Y} \triangleq (\Phi, \Omega)$$

where $\Phi$ is a binary random variable which is affected by $X$, and $\Omega$ is a non-negative random variable which represents the absolute value of the LLR. For each index $i \in \mathcal{I}(j)$, the *pdf* of $\Omega_i$ is independent of $i$, and it gets the form

$$f_{\Omega_i}(\omega) \triangleq f_{\Omega}(\omega; j) = \begin{cases} a(\omega; j) + a(-\omega; j) = (1 + e^{-\omega})\, a(\omega; j) & \text{if} \quad \omega > 0 \\ a(0; j) & \text{if} \quad \omega = 0 \end{cases}. \tag{9}$$

We note that the transition in case $\omega > 0$ follows from the symmetry property of $a(\cdot; j)$, and that the random variable $\Omega_i$ is statistically independent of $X_i$ (where $i \in \{1, \ldots, n\}$).

Denoting by $R$ the rate of the code $\mathcal{C}$, since the codewords are transmitted with equal probability

$$H(\mathbf{X}) = nR. \tag{10}$$

Also, since the $J$ parallel channels are memoryless, then

$$H(\mathbf{Y}|\mathbf{X}) = \sum_{i=1}^{n} H(Y_i|X_i). \tag{11}$$

[1] In case the output alphabets of the component channels are not equal, then $\mathcal{Y}$ can be defined as their union.

The mapping $Y_i \to \widetilde{Y}_i$ is memoryless, hence $H(\widetilde{\mathbf{Y}}|\mathbf{Y}) = \sum_{i=1}^{n} H(\widetilde{Y}_i|Y_i)$, and

$$\begin{aligned} H(\mathbf{Y}) &= H(\widetilde{\mathbf{Y}}) - H(\widetilde{\mathbf{Y}}|\mathbf{Y}) + H(\mathbf{Y}|\widetilde{\mathbf{Y}}) \\ &= H(\widetilde{\mathbf{Y}}) - \sum_{i=1}^{n} H(\widetilde{Y}_i|Y_i) + H(\mathbf{Y}|\widetilde{\mathbf{Y}}) \end{aligned} \tag{12}$$

$$\begin{aligned} H(\mathbf{Y}|\widetilde{\mathbf{Y}}) &\leq \sum_{i=1}^{n} H(Y_i|\widetilde{Y}_i) \\ &= \sum_{i=1}^{n} \Big[ H(Y_i) - H(\widetilde{Y}_i) + H(\widetilde{Y}_i|Y) \Big] \,. \end{aligned} \tag{13}$$

Applying the above towards the derivation of a lower bound on the conditional entropy $H(\mathbf{X}|\mathbf{Y})$, we get

$$\begin{aligned} H(\mathbf{X}|\mathbf{Y}) &= H(\mathbf{X}) + H(\mathbf{Y}|\mathbf{X}) - H(\mathbf{Y}) \\ &\stackrel{(a)}{=} nR + \sum_{i=1}^{n} H(Y_i|X_i) - H(\widetilde{\mathbf{Y}}) - H(\mathbf{Y}|\widetilde{\mathbf{Y}}) + \sum_{i=1}^{n} H(\widetilde{Y}_i|Y_i) \\ &\stackrel{(b)}{\geq} nR + \sum_{i=1}^{n} H(Y_i|X_i) - H(\widetilde{\mathbf{Y}}) - \sum_{i=1}^{n} \Big[ H(Y_i) - H(\widetilde{Y}_i) + H(\widetilde{Y}_i|Y_i) \Big] + \sum_{i=1}^{n} H(\widetilde{Y}_i|Y_i) \\ &= nR - H(\widetilde{\mathbf{Y}}) + \sum_{i=1}^{n} H(\widetilde{Y}_i) - \sum_{i=1}^{n} \big[ H(Y_i) - H(Y_i|X_i) \big] \\ &= nR - H(\widetilde{\mathbf{Y}}) + \sum_{i=1}^{n} H(\widetilde{Y}_i) - \sum_{i=1}^{n} I(X_i;Y_i) \\ &\stackrel{(c)}{\geq} nR - H(\widetilde{\mathbf{Y}}) + \sum_{i=1}^{n} H(\widetilde{Y}_i) - \sum_{j=1}^{J} n^{[j]}\, C_j \end{aligned} \tag{14}$$

where (a) relies on (10)–(12), (b) relies on (13), and (c) follows since $I(X_i;Y_i) \leq C_j$ for all $i \in \mathcal{I}(j)$, and $|\mathcal{I}(j)| = n^{[j]}$ for $j \in \{1, \ldots, J\}$. In order to obtain a lower bound on $H(\mathbf{X}|\mathbf{Y})$ from (14), we will calculate the exact entropy of the random variables $\{\widetilde{Y}_i\}$, and find an upper bound on the entropy of the random vector $\widetilde{\mathbf{Y}}$. This will finally provide the lower bound on the conditional entropy given in (7). Considering an index $i \in \mathcal{I}(j)$ for some $j \in \{1, 2, \ldots J\}$, we get

$$\begin{aligned} H(\widetilde{Y}_i) &= H(\Phi_i, \Omega_i) \\ &= H(\Omega_i) + H(\Phi_i|\Omega_i) \\ &= H(\Omega_i) + E_\omega \left[ H(\Phi_i|\Omega_i = \omega) \right] \\ &= H(\Omega_i) + 1 \end{aligned} \tag{15}$$

where the last transition is due to the fact that given the absolute value of the LLR, since the parallel channels are MBIOS and the coded bits are equally likely to be 0 or 1, the sign of the LLR is equally likely to be positive or negative. The entropy $H(\Omega_i)$ is not expressed explicitly as it cancels out later. In the following, we derive an upper bound on $H(\widetilde{\mathbf{Y}})$:

$$\begin{aligned} H(\widetilde{\mathbf{Y}}) &= H\big((\Phi_1, \ldots, \Phi_n),\ (\Omega_1, \ldots, \Omega_n)\big) \\ &= H(\Omega_1, \ldots, \Omega_n) + H\big((\Phi_1, \ldots, \Phi_n) \mid (\Omega_1, \ldots, \Omega_n)\big) \\ &= \sum_{i=1}^{n} H(\Omega_i) + H\big((\Phi_1, \ldots, \Phi_n) \mid (\Omega_1, \ldots, \Omega_n)\big) \end{aligned} \tag{16}$$

where the last equality follows since the random variables $\Omega_i$ are statistically independent.

Let us introduce the assignment $f : \{+1, -1\} \to \{0, 1\}$ which maps back the $+1$ and $-1$ to 0 and 1, respectively, and let $\widetilde{\Phi}_i \triangleq f(\Phi_i)$. Since $\Phi_i = \Theta_i X_i$, then

$$\widetilde{\Phi}_i = \widetilde{\Theta}_i + \widetilde{X}_i, \quad i = 1, \ldots, n$$

where the last addition is modulo-2. Define the syndrome vector

$$\mathbf{S} = (\widetilde{\Phi}_1, \ldots, \widetilde{\Phi}_n) H^T$$

where $H$ is a $c \times n$ parity-check matrix of the binary linear block code $\mathcal{C}$, and let $L$ be the index of the vector $(\widetilde{\Phi}_1, \ldots, \widetilde{\Phi}_n)$ in the coset which corresponds to $\mathbf{S}$. Since each coset has exactly $2^{nR}$ elements which are equally likely then $H(L) = nR$, and we get

$$\begin{aligned} H\big((\Phi_1, \ldots, \Phi_n) \mid (\Omega_1, \ldots, \Omega_n)\big) &= H\big(\mathbf{S}, L \mid (\Omega_1, \ldots, \Omega_n)\big) \\ &\leq H(L) + H\big(\mathbf{S} \mid (\Omega_1, \ldots, \Omega_n)\big) \\ &= nR + H\big(\mathbf{S} \mid (\Omega_1, \ldots, \Omega_n)\big) \\ &\leq nR + \sum_{m=1}^{c} H\big(S_m \mid (\Omega_1, \ldots, \Omega_n)\big). \end{aligned} \tag{17}$$

Since $(\widetilde{x}_1, \ldots, \widetilde{x}_n)$ denotes a selected transmitted codeword in $\mathcal{C}$ (i.e., before converting its symbols to $\pm 1$ and their transmission over the parallel channels), then

$$\begin{aligned} \mathbf{S} &= (\widetilde{\Phi}_1, \ldots, \widetilde{\Phi}_n) H^T \\ &= (\widetilde{\Theta}_1, \ldots, \widetilde{\Theta}_n) H^T + (\widetilde{X}_1, \ldots, \widetilde{X}_n) H^T \\ &= (\widetilde{\Theta}_1, \ldots, \widetilde{\Theta}_n) H^T. \end{aligned}$$

Let us look at the $m$-th parity-check equation which involves $k_m$ variables, and assume that the set of indices of these variables is $\{i_1, \ldots, i_{k_m}\}$. Then, the component $S_m$ of the syndrome is equal to 1 if and only if there is an odd number of ones in the random vector $(\widetilde{\Theta}_{i_1}, \ldots, \widetilde{\Theta}_{i_{k_m}})$. To calculate the probability that $S_m$ is equal to 1, we rely on the following lemma:

**Lemma 3.1 ([17], Lemma 4.1).** If the $m$-th linear constraint defined by the parity-check matrix $H$ involves $k_m$ variables, and if $\{i_1, \ldots, i_{k_m}\}$ denote the indices of these variables, then

$$\Pr\left(S_m = 1 \mid (\Omega_{i_1}, \ldots, \Omega_{i_{k_m}}) = (\alpha_1, \ldots, \alpha_{k_m})\right) = \frac{1}{2}\left[1 - \prod_{w=1}^{k_m} \tanh\left(\frac{\alpha_w}{2}\right)\right]. \tag{18}$$

From this lemma, we obtain

$$H\left(S_m | (\Omega_{i_1}, \ldots, \Omega_{i_{k_m}}) = (\alpha_1, \ldots, \alpha_{k_m})\right) = h_2\left(\frac{1}{2}\left[1 - \prod_{w=1}^{k_m} \tanh\left(\frac{\alpha_w}{2}\right)\right]\right)$$

where $h_2(\cdot)$ denotes the binary entropy function to base 2. By taking the statistical expectation over the $k_m$ random variables $\Omega_{i_1}, \ldots, \Omega_{i_{k_m}}$, we get

$$\begin{aligned} &H\left(S_m | (\Omega_{i_1}, \ldots, \Omega_{i_{k_m}})\right) \\ &= \int_0^\infty \ldots \int_0^\infty h_2\left(\frac{1}{2}\left[1 - \prod_{w=1}^{k_m} \tanh\left(\frac{\alpha_w}{2}\right)\right]\right) \prod_{w=1}^{k_m} f_{\Omega_{i_w}}(\alpha_w)\, d\alpha_1 d\alpha_2 \ldots d\alpha_{k_m}\,. \end{aligned}$$

Let $\beta_{j,m}$ denote the number of indices $w \in \{i_1, \ldots, i_{k_m}\}$ referring to variables which are transmitted over the $j^{\text{th}}$ channel. From the Taylor series expansion of $h_2(\cdot)$ around $\frac{1}{2}$ (see [17, Appendix C.2])

$$h_2(x) = 1 - \frac{1}{2\ln 2} \sum_{p=1}^{\infty} \frac{(1-2x)^{2p}}{p(2p-1)} \,, \quad 0 \leq x \leq 1 \tag{19}$$

it follows [17, Appendix C.3] that

$$\begin{aligned} & H\left(S_m | (\Omega_{i_1}, \ldots, \Omega_{i_{k_m}})\right) \\ & = 1 - \frac{1}{2\ln 2} \sum_{p=1}^{\infty} \left\{ \frac{1}{p(2p-1)} \prod_{w=1}^{k_m} \left( \int_0^{\infty} f_{\Omega_{i_w}}(\alpha) \, \tanh^{2p}\left(\frac{\alpha}{2}\right) \, d\alpha \right) \right\} \\ & = 1 - \frac{1}{2\ln 2} \sum_{p=1}^{\infty} \left\{ \frac{1}{p(2p-1)} \prod_{j=1}^{J} \left( \int_0^{\infty} f_{\Omega}(\alpha; j) \, \tanh^{2p}\left(\frac{\alpha}{2}\right) \, d\alpha \right)^{\beta_{j,m}} \right\} \end{aligned} \tag{20}$$

where the first transition is based on (19) and goes along the same lines as [17, Appendix C.3]), and the second transition is due to the fact that for all $i \in \mathcal{I}(j)$, the *pdf* of the random variable $\Omega_i$ is independent of $i$, see (9). Summing over all the parity-check equations of $H$ gives

$$\begin{aligned} & \sum_{m=1}^{c} H\big(S_m | (\Omega_1, \ldots, \Omega_n)\big) \\ & = c - \frac{1}{2\ln 2} \sum_{p=1}^{\infty} \frac{1}{p(2p-1)} \sum_{m=1}^{c} \left\{ \prod_{j=1}^{J} \left( \int_0^{\infty} f_{\Omega}(\alpha; j) \, \tanh^{2p}\left(\frac{\alpha}{2}\right) \, d\alpha \right)^{\beta_{j,m}} \right\}. \end{aligned} \tag{21}$$

By combining (9), (16), (17) and (21), we get the following upper bound on $H(\widetilde{\mathbf{Y}})$:

$$\begin{aligned} H(\widetilde{\mathbf{Y}}) \leq & \sum_{i=1}^{n} H(\Omega_i) + nR \\ & + c \left[ 1 - \frac{1}{2c\ln 2} \sum_{p=1}^{\infty} \frac{1}{p(2p-1)} \sum_{m=1}^{c} \left\{ \prod_{j=1}^{J} \left( \int_0^{\infty} a(\alpha; j)(1 + e^{-\alpha}) \tanh^{2p}\left(\frac{\alpha}{2}\right) d\alpha \right)^{\beta_{j,m}} \right\} \right] \\ \stackrel{(a)}{=} & \sum_{i=1}^{n} H(\Omega_i) + nR + n(1 - R_{\text{d}}) \left[ 1 - \frac{1}{2n(1-R_{\text{d}})\ln 2} \sum_{p=1}^{\infty} \left\{ \frac{1}{p(2p-1)} \sum_{m=1}^{c} \prod_{j=1}^{J} {g_{j,p}}^{\beta_{j,m}} \right\} \right] \end{aligned} \tag{22}$$

where $(a)$ relies on the definition of $g_{j,p}$ in (8) and since $R_{\text{d}} \triangleq 1 - \frac{c}{n}$ denotes the design rate of $\mathcal{C}$. Finally, the substitution of (15) and (22) in the RHS of (14) provides the lower bound on the conditional entropy $H(\mathbf{X}|\mathbf{Y})$ given in (7). This completes the proof of the proposition. □

## 3.2 Upper Bound on the Conditional Entropy

In this section, we provide an upper bound on the conditional entropy of a transmitted codeword given the received sequence. The bound holds for an arbitrary binary linear block code whose transmission takes place over a set of parallel channels, and is expressed in terms of the code rate and the bit-error probability of the code which is associated with an arbitrary decoder.

**Lemma 3.2.** Let $\mathcal{C}$ be a binary linear block code of length $n$ and rate $R$, and assume that its transmission takes place over a set of parallel channels. Let $\mathbf{x} = (x_1, \ldots, x_n)$ and $\mathbf{y} = (y_1, \ldots, y_n)$ designate the transmitted codeword and the received sequence, respectively. Then

$$\frac{H(\mathbf{X}|\mathbf{Y})}{n} \leq R\, h_2(P_{\mathrm{b}}) \tag{23}$$

where $P_{\mathrm{b}}$ designates the bit error probability of the code $\mathcal{C}$.

*Proof.* Since there is a one to one correspondence between the codewords and the set of information bits used to encode them, then $H(\mathbf{X}|\mathbf{Y}) = H(\mathbf{U}|\mathbf{Y})$ where the vector $\mathbf{u} = (u_1, \ldots, u_{nR})$ denotes the sequence of information bits used to encode the codeword $\mathbf{x}$. Let $P_{\mathrm{b}}^{(i)}(\mathcal{C})$ denote the probability of decoding the bit $u_i$ erroneously given the received sequence at the output of the set of parallel channels, then the bit error probability is given by

$$P_{\mathrm{b}}(\mathcal{C}) = \frac{1}{nR} \sum_{i=1}^{nR} P_{\mathrm{b}}^{(i)}(\mathcal{C}). \tag{24}$$

This therefore gives

$$\begin{aligned}
\frac{H(\mathbf{X}|\mathbf{Y})}{n} &= \frac{H(\mathbf{U}|\mathbf{Y})}{n} \\
&\stackrel{(a)}{\leq} \frac{1}{n} \sum_{i=1}^{nR} H(U_i|\mathbf{Y}) \\
&\stackrel{(b)}{\leq} \frac{1}{n} \sum_{i=1}^{nR} h_2\big(P_{\mathrm{b}}^{(i)}(\mathcal{C})\big) \\
&\stackrel{(c)}{\leq} R\, h_2\left(\frac{1}{nR} \sum_{i=1}^{nR} P_{\mathrm{b}}^{(i)}(\mathcal{C})\right) \\
&\stackrel{(d)}{=} R\, h_2\big(P_{\mathrm{b}}(\mathcal{C})\big)
\end{aligned}$$

where inequality (a) holds from the chain rule of the entropy and since conditioning reduces entropy, inequality (b) follows from Fano's inequality and since the code is binary, inequality (c) is based on Jensen's inequality and the concavity of $h_2(\cdot)$, and equality (d) follows from (24). □

# 4 An Upper bound on the Achievable Rates of LDPC codes over Parallel Channels

In this section, we derive an upper bound on the design rate of a sequence of ensembles of LDPC codes whose transmission takes place over a set of statistically independent parallel MBIOS channels, and achieves vanishing bit error probability under ML decoding. This bound is used in the next section for the derivation of an upper bound on the design rate of an arbitrary sequence of ensembles of punctured LDPC codes.

Let us assume that a binary LDPC code $\mathcal{C}$ of length $n$ is transmitted over a set of $J$ statistically independent parallel MBIOS channels. Denote the number of code bits of $\mathcal{C}$ which are transmitted over the $j^{\text{th}}$ channel by $n^{[j]}$, and the fraction of bits transmitted over the $j^{\text{th}}$ channel by

$$p_j \triangleq \frac{n^{[j]}}{n}, \quad j \in \{1, \ldots, J\}. \tag{25}$$

Let $\mathcal{G}$ be a bipartite graph which represents the code $\mathcal{C}$, and $E$ be the set of edges in $\mathcal{G}$. Let $E^{[j]}$ designate the set of edges connected to variable nodes which correspond to code bits transmitted over the $j^{\text{th}}$ channel, and

$$q_j \triangleq \frac{|E^{[j]}|}{|E|}, \quad j \in \{1, \ldots, J\} \tag{26}$$

denote the fraction of edges connected to these variable nodes. Referring to the edges from the subset $E^{[j]}$, let $\lambda_i^{[j]}$ designate the fraction of these edges which are connected to variable nodes of degree $i$, and define the following $J$ degree distributions from the edge perspective:

$$\lambda^{[j]}(x) \triangleq \sum_{i=2}^{\infty} \lambda_i^{[j]} x^{i-1} , \qquad j \in \{1, \ldots, J\}$$

which correspond to each of the $J$ parallel channels. According to this notation, the number of edges connected to variable nodes corresponding to code bits transmitted over the $j^{\text{th}}$ channel is given by

$$|E^{[j]}| = \frac{n^{[j]}}{\displaystyle\sum_{i=2}^{\infty} \frac{\lambda_i^{[j]}}{i}} , \qquad j \in \{1, \ldots, J\}. \tag{27}$$

For the simplicity of the notation, let us define a vector of degree distributions for the variable nodes from the edge perspective to be $\underline{\lambda}(x) = \big(\lambda^{[1]}(x), \ldots, \lambda^{[J]}(x)\big)$. Following the notation in [11], the ensemble $(n, \underline{\lambda}, \rho)$ is defined to be the set of LDPC codes of length $n$, which according to their representation by bipartite graphs and the assignment of their code bits to the parallel channels, imply left and right degree distributions of $\underline{\lambda}$ and $\rho$, respectively.

**Lemma 4.1.**

$$\frac{1}{\int_0^1 \lambda(x)\, dx} = \sum_{j=1}^{J} \left\{ \frac{p_j}{\int_0^1 \lambda^{[j]}(x)\, dx} \right\}. \tag{28}$$

where $\lambda$ is the original left degree distribution which serves to construct the vector of left degree distributions $\underline{\lambda}$, due to the transmission of the LDPC code over parallel channels.

*Proof.* Since $E^{[1]}, \ldots, E^{[J]}$ forms a sequence of disjoint sets whose union is the set $E$, we get the equality $|E| = \sum_{j=1}^{J} |E^{[j]}|$. From (27), we therefore get

$$\frac{n}{\sum_{i=2}^{\infty} \frac{\lambda_i}{i}} = \sum_{j=1}^{J} \left\{ \frac{n^{[j]}}{\sum_{i=2}^{\infty} \frac{\lambda_i^{[j]}}{i}} \right\} \tag{29}$$

and by dividing both sides of the equality by $n$ and using (25), the lemma follows immediately. □

**Lemma 4.2.**

$$q_j = \frac{p_j}{\int_0^1 \lambda^{[j]}(x)\, dx} \cdot \frac{1}{\displaystyle\sum_{k=1}^{J} \left\{ \frac{p_k}{\int_0^1 \lambda^{[k]}(x)\, dx} \right\}}, \quad \forall\, j \in \{1, \ldots, J\}. \tag{30}$$

*Proof.* The lemma follows directly from (26), (27) and Lemma 4.1. □

In the following, we introduce a sequence of ensembles of LDPC codes, say $\{(n_r, \underline{\lambda}_r, \rho)\}_{r=1}^{\infty}$ where the fraction of code bits assigned to each of the $J$ parallel channels is uniform over all the codes of each ensemble, and $\rho$ is fixed for all the ensembles of this sequence (i.e., it is independent of $r$). Since $\lambda$ which corresponds to the overall left degree distribution of the edges is independent of $r$ (and so is the right degree distribution $\rho$), one can consider here the common design rate of the sequence of ensembles $\{(n_r, \underline{\lambda}_r, \rho)\}_{r=1}^{\infty}$ which does not depend on $r$.

This setting is general enough for applying the following theorem to various applications which form particular cases of communication over parallel channels, e.g., punctured LDPC codes [2, 4], non-uniformly error protected LDPC codes [11], and LDPC-coded modulation (see e.g., [6, 16]). In this setting, the fraction of code bits assigned to the $j^{\text{th}}$ channel, $p_{j,r}$, depends on $j \in \{1, \ldots, J\}$ and $r \in \mathbb{N}$, but not on the particular code chosen from each ensemble. It follows from Lemma 4.2 that the same property also holds for $q_{j,r}$ which designates the fraction of edges which are connected to variable nodes whose code bits are assigned to the $j^{\text{th}}$ channel. In the following, we assume that the limits

$$p_j \triangleq \lim_{r \to \infty} p_{j,r}, \quad q_j \triangleq \lim_{r \to \infty} q_{j,r} \tag{31}$$

exist and are positive for all $j \in \{1, \ldots, J\}$.

**Theorem 4.1.** Let a sequence of LDPC ensembles $\{(n_r, \underline{\lambda}_r, \rho)\}_{r=1}^{\infty}$ be transmitted over a set of $J$ statistically independent parallel MBIOS channels, and assume that the block length $(n_r)$ goes to infinity as we let $r$ tend to infinity. Let $C_j$ denote the capacity of the $j^{\text{th}}$ channel, and $a(\cdot; j)$ designate the *pdf* of the LLR at the output of the $j^{\text{th}}$ channel given its input is 1. If in the limit where $r$ tends to infinity, the bit error probability of this sequence under ML decoding vanishes, then the common design rate $R_{\text{d}}$ of these ensembles satisfies

$$R_{\text{d}} \leq 1 - \frac{1 - \displaystyle\sum_{j=1}^{J} p_j C_j}{1 - \displaystyle\frac{1}{2 \ln 2} \sum_{p=1}^{\infty} \left\{ \frac{1}{p(2p-1)} \, \Gamma\left( \sum_{j=1}^{J} q_j \, g_{j,p} \right) \right\}} \tag{32}$$

where $\Gamma$, as introduced in (3), denotes the right degree distribution from the node perspective, and $g_{j,p}$ is introduced in (8).

*Proof.* Let $\{\mathcal{C}_r\}_{r=1}^{\infty}$ be a sequence of LDPC codes chosen uniformly at random from the sequence of ensembles $\{(n_r, \underline{\lambda}_r, \rho)\}_{r=1}^{\infty}$. Denote the rate of the code $\mathcal{C}_r$ by $R_r$, and let $P_{\text{b},r}$ be its bit error probability under ML decoding. Let $\mathcal{G}_r$ be a bipartite graph of the code $\mathcal{C}_r$ whose left and right degree distributions from the edge perspective are $\underline{\lambda}_r$ and $\rho$, respectively. From Proposition 3.1 and Lemma 3.2, it follows that the following inequality holds for the binary linear block code $\mathcal{C}_r$:

$$\begin{aligned} R_r h(P_{\text{b},r}) \quad \geq \quad & 1 - \sum_{j=1}^{J} p_{j,r} C_j \\ & -(1 - R_{\text{d}}) \left( 1 - \frac{1}{2 n_r (1 - R_{\text{d}}) \ln 2} \sum_{p=1}^{\infty} \left\{ \frac{1}{p(2p-1)} \sum_{m=1}^{n_r(1-R_{\text{d}})} \prod_{j=1}^{J} (g_{j,p})^{\beta_{r,j,m}} \right\} \right) \end{aligned} \tag{33}$$

where $n_r$ is the block length of the code $\mathcal{C}_r$, $R_{\text{d}}$ is the asymptotic design rate for all the codes from the sequence of ensembles $\{(n_r, \underline{\lambda}_r, \rho)\}_{r=1}^{\infty}$, and $\beta_{r,j,m}$ denotes the number of edges which are connected to the $m^{\text{th}}$ parity-check node of the graph $\mathcal{G}_r$ and are related to code bits transmitted over the $j^{\text{th}}$ channel (where $j \in \{1, \ldots, J\}$ and $m \in \{1, \ldots n_r(1-R_{\text{d}})\}$). By taking the expectation on both sides of (33) and letting $r$ tend to infinity, we get

$$\begin{aligned} & 1 - \sum_{j=1}^{J} p_j C_j - (1-R_{\text{d}}) \\ & \qquad \cdot \lim_{r\to\infty} \left( 1 - \frac{1}{2n_r(1-R_{\text{d}}) \ln 2} \sum_{p=1}^{\infty} \left\{ \frac{1}{p(2p-1)} \sum_{m=1}^{n_r(1-R_{\text{d}})} \mathbb{E}\left( \prod_{j=1}^{J} (g_{j,p})^{\beta_{r,j,m}} \right) \right\} \right) \leq 0. \qquad (34) \end{aligned}$$

The last result follows from the LHS of (33), due to the concavity of the binary entropy function and Jensen's inequality, and since by our assumption, the bit error probability of the ensembles vanishes in the limit where $r$ tends to infinity.

The derivation of an upper bound on the design rate is proceeded by calculating the expectation of the product inside the LHS of (34). Let $k_{r,m}$ denote the degree of the $m^{\text{th}}$ parity-check node of the bipartite graph $\mathcal{G}_r$, then the smoothing theorem gives

$$\mathbb{E}\left( \prod_{j=1}^{J} (g_{j,p})^{\beta_{r,j,m}} \right) = \mathbb{E}\left[ \mathbb{E}\left( \prod_{j=1}^{J} (g_{j,p})^{\beta_{r,j,m}} \;\Big|\; \sum_{j=1}^{J} \beta_{r,j,m} = k_{r,m} \right) \right] \qquad (35)$$

where the outer expectation is carried over the random variable $k_{r,m}$. We first calculate the inner expectation in the RHS of (35). It follows from (27) that the number of edges, $|E_r^{[j]}| \triangleq |E^{[j]}(\mathcal{G}_r)|$, connected to variable nodes corresponding to code bits transmitted over the $j^{\text{th}}$ channel, is independent of the code $\mathcal{C}_r$ we choose from the ensemble $(n_r, \underline{\lambda}, \rho)$. The same property also holds for the total number of edges in the graph (since $|E_r| = \sum_{j=1}^{J} |E_r^{[j]}|$). Since the code $\mathcal{C}_r$ is chosen uniformly at random from the ensemble, it follows that if $k_{r,m}$ is a given positive integer, then

$$\begin{aligned} & \mathbb{E}\left( \prod_{j=1}^{J} (g_{j,p})^{\beta_{r,j,m}} \;\Big|\; \sum_{j=1}^{J} \beta_{r,j,m} = k_{r,m} \right) \\ & = \sum_{\substack{b_1, \ldots, b_J \geq 0 \\ \sum_{j=1}^{J} b_j = k_{r,m}}} \left\{ \Pr\big(\beta_{r,j,m} = b_j \;,\; \forall j \in \{1, \ldots, J\}\big) \prod_{j=1}^{J} (g_{j,p})^{b_j} \right\} \\ & = \sum_{\substack{b_1, \ldots, b_J \geq 0 \\ \sum_{j=1}^{J} b_j = k_{r,m}}} \left\{ \frac{\binom{|E_r^{[1]}|}{b_1} \cdot \ldots \cdot \binom{|E_r^{[J]}|}{b_J}}{\binom{|E_r|}{k_{r,m}}} \prod_{j=1}^{J} (g_{j,p})^{b_j} \right\}. \qquad (36) \end{aligned}$$

**Lemma 4.3.**

$$\lim_{r\to\infty} \frac{\binom{|E_r^{[1]}|}{b_1} \cdot \ldots \cdot \binom{|E_r^{[J]}|}{b_J}}{\binom{|E_r|}{k_{r,m}}} = \prod_{j=1}^{J} (q_j)^{b_j} \lim_{r\to\infty} \binom{k_{r,m}}{b_1, b_2, \ldots, b_J}. \qquad (37)$$

*Proof.* By assumption, in the limit where we let $r$ tend to infinity, the block length $n_r$ also tends to infinity. Hence, from (27), we get that for all $j \in \{1, \ldots, J\}$, also $E_r^{[j]}$ tends to infinity in the limit where $r$ tends to infinity.

$$
\begin{aligned}
&\lim_{r\to\infty} \frac{\binom{|E_r^{[1]}|}{b_1}\cdot \ldots \cdot \binom{|E_r^{[J]}|}{b_J}}{\binom{|E_r|}{k_{r,m}}} \\
&= \lim_{r\to\infty} \frac{|E_r^{[1]}|! \ldots |E_r^{[J]}|!}{|E_r|!} \, \frac{(|E_r|-k_{r,m})!}{(|E_r^{[1]}|-b_1)! \ldots (|E_r^{[J]}|-b_J)!} \binom{k_{r,m}}{b_1, b_2, \ldots, b_J} \\
&= \lim_{r\to\infty} \left\{ \frac{|E_r^{[1]}|! \ldots |E_r^{[J]}|!}{|E_r|!} \, \frac{(|E_r|-k_{r,m})!}{(|E_r^{[1]}|-b_1)! \ldots (|E_r^{[J]}|-b_J)!} \right\} \lim_{r\to\infty} \binom{k_{r,m}}{b_1, b_2, \ldots, b_J} \\
&\overset{(a)}{=} \lim_{r\to\infty} \frac{|E_r^{[1]}|^{b_1} \ldots |E_r^{[J]}|^{b_J}}{|E_r|^{k_{r,m}}} \lim_{r\to\infty} \binom{k_{r,m}}{b_1, b_2, \ldots, b_J} \\
&\overset{(b)}{=} \lim_{r\to\infty} \left(\frac{|E_r^{[1]}|}{|E_r|}\right)^{b_1} \cdots \left(\frac{|E_r^{[J]}|}{|E_r|}\right)^{b_J} \lim_{r\to\infty} \binom{k_{r,m}}{b_1, b_2, \ldots, b_J} \\
&\overset{(c)}{=} \lim_{r\to\infty} \prod_{j=1}^{J} (q_{j,r})^{b_j} \lim_{r\to\infty} \binom{k_{r,m}}{b_1, b_2, \ldots, b_J} \\
&= \prod_{j=1}^{J} (q_j)^{b_j} \lim_{r\to\infty} \binom{k_{r,m}}{b_1, b_2, \ldots, b_J}
\end{aligned}
$$

where equality (a) follows since for all $j \in \{1, \ldots, J\}$, $|E_r^{[j]}| \to \infty$ as we let $r$ tend to infinity, while on the other hand, the maximal right degree (and hence, also $b_1, \ldots, b_J$ and $k_{r,m}$) stay bounded; equality (b) is valid due to the constraint $\sum_{j=1}^{J} b_j = k_{r,m}$, and equality (c) follows from (26). $\square$

By letting $r$ tend to infinity in both sides of (35), and substituting (36) and (37) in the RHS of (35), we get that for all $p \in \mathbb{N}$

$$
\begin{aligned}
&\lim_{r\to\infty} \mathbb{E}\left[\prod_{j=1}^{J} (g_{j,p})^{\beta_{r,j,m}}\right] \\
&\overset{(a)}{=} \mathbb{E}\left[\lim_{r\to\infty} \mathbb{E}\left(\prod_{j=1}^{J} (g_{j,p})^{\beta_{r,j,m}} \;\Big|\; \sum_{j=1}^{J} \beta_{r,j,m} = k_{r,m}\right)\right] \\
&\overset{(b)}{=} \mathbb{E}\Bigg[\lim_{r\to\infty} \sum_{\substack{b_1,\ldots,b_J \ge 0 \\ \sum_{j=1}^{J} b_j = k_{r,m}}} \binom{k_{r,m}}{b_1, b_2, \ldots, b_J} \prod_{j=1}^{J} (q_j \; g_{j,p})^{b_j}\Bigg] \\
&= \mathbb{E}\Bigg[\lim_{r\to\infty} \bigg(\sum_{j=1}^{J} q_j g_{j,p}\bigg)^{k_{r,m}}\Bigg] \\
&\overset{(c)}{=} \sum_{k=1}^{d_{c,\max}} \left\{\Gamma_k \,\Big(\sum_{j=1}^{J} q_j g_{j,p}\Big)^k\right\} \\
&= \Gamma\bigg(\sum_{j=1}^{J} q_j g_{j,p}\bigg)
\end{aligned}
\tag{38}
$$

where equality (a) follows from (35) and since the right degree distribution is independent of $r$ (note that the outer expectation in equality (a) is performed w.r.t. the degree of the $m^{\text{th}}$ parity-check node); equality (b) follows from (36) and (37), and since the number of terms in the sum is bounded (this number is upper bounded by $(k_{r,m})^{J-1}$, so it is bounded for all $r \in \mathbb{N}$ due to the fact that the maximal right degree is fixed), and equality (c) follows since the right degree distribution is independent of $r$. Since the limit in (38) does not depend on the index $m$ which appears in the inner summation at the LHS of (34) and also $\lim_{r\to\infty} n_r(1-R_{\text{d}}) = \infty$, then we get from (38)

$$\begin{aligned} &\lim_{r\to\infty} \frac{1}{n_r(1-R_{\text{d}})} \sum_{m=1}^{n_r(1-R_{\text{d}})} \mathbb{E}\left[\prod_{j=1}^{J} (g_{j,p})^{\beta_{r,j,m}}\right] \\ &\stackrel{\text{(a)}}{=} \lim_{r\to\infty} \mathbb{E}\left[\prod_{j=1}^{J} (g_{j,p})^{\beta_{r,j,m}}\right] \\ &\stackrel{\text{(b)}}{=} \Gamma\left(\sum_{j=1}^{J} q_j g_{j,p}\right) \end{aligned} \tag{39}$$

where equality (a) follows from the fact that if $\{a_r\}$ is a convergent sequence then the equality $\lim_{r\to\infty} \frac{1}{r}\sum_{i=1}^{r} a_i = \lim_{r\to\infty} a_r$ holds, and also since any sub-sequence of a convergent sequence converges to the same limit as of the original sequence; equality (b) follows from (38). Combining (34) and (39) gives

$$1 - \sum_{j=1}^{J} p_j C_j - (1-R_{\text{d}})\left(1 - \frac{1}{2\ln 2}\sum_{p=1}^{\infty}\left\{\frac{1}{p(2p-1)}\,\Gamma\left(\sum_{j=1}^{J} q_j g_{j,p}\right)\right\}\right) \leq 0.$$

Finally, solving the last inequality for $R_{\text{d}}$ gives the upper bound on the design rate in (32). $\square$

**Example 4.1.** For the particular case where the $J$ MBIOS parallel channels are binary erasure channels where the erasure probability of the $j^{\text{th}}$ channel is $\varepsilon_j$, we get from (8)

$$g_{j,p} = 1 - \varepsilon_j\ , \quad \forall\, j \in \{1,\ldots,J\},\ \ p \in \mathbb{N}. \tag{40}$$

Since $g_{j,p}$ is independent of $p$ for a BEC, and based on the equality $\sum_{p=1}^{\infty} \frac{1}{2p(2p-1)} = \ln 2$, we obtain from Theorem 4.1 that the common design rate of the sequence of LDPC ensembles is upper bounded by

$$R_{\text{d}} \leq 1 - \frac{\sum_{j=1}^{J} p_j \varepsilon_j}{1 - \Gamma\left(1 - \sum_{j=1}^{J} q_j\, \varepsilon_j\right)}\,. \tag{41}$$

This particular result coincides with [11, Theorem 2].

The proof of Theorem 4.1 relies on the assumption that the right degree distribution $\rho$ is fixed, and does not depend on the ordinal number $r$ of the ensemble. For a capacity-achieving sequence of LDPC ensembles, both the maximal and the average right degrees tend to infinity (see [15, Theorem 1]). Hence, for a capacity-achieving sequence of LDPC codes, $\rho$ cannot be fixed.

**Remark 4.1.** We wish to discuss a possible refinement of the statement in Theorem 4.1. Let us assume that the (overall) degree distributions $\lambda$ and $\rho$ are fixed, but due to the transmission over parallel channels, the corresponding vector of degree distributions $\underline{\lambda}_r = (\lambda_r^{[1]}, \ldots, \lambda_r^{[J]})$ and also $p_{j,r}$ and $q_{j,r}$ depend on the code from the ensemble $(n_r, \lambda, \rho)$. Since the derivation of this theorem relies on the bounds on the conditional entropy from Section 3 (which are valid code by code), one can refine the statement in Theorem 4.1 so that the modified theorem permits the dependency of the vector $(\lambda_r^{[1]}, \ldots, \lambda_r^{[J]})$ on the specific code chosen from the ensemble. In this case, the equalities in (31) are transformed to

$$p_j = \lim_{r\to\infty} \mathbb{E}\left[p_{j,r}(\mathcal{C})\right], \quad q_j = \lim_{r\to\infty} \mathbb{E}\left[q_{j,r}(\mathcal{C})\right]$$

where the expectation is carried over the code $\mathcal{C}$ from the ensemble $(n_r, \lambda, \rho)$. In this case, the proof of Theorem 4.1 involves an expectation over $\mathcal{C}$ on both sides of (33) (which is valid code by code) and then we let $r$ tend to infinity, as in (34). By invoking Jensen's inequality, Lemma 4.3 is changed under the above assumption to the inequality

$$\lim_{r\to\infty} \mathbb{E}_{\mathcal{C}}\left[\frac{\binom{|E_r^{[1]}|}{b_1}\cdot\ldots\cdot\binom{|E_r^{[J]}|}{b_J}}{\binom{|E_r|}{k_{r,m}}}\right] \geq \prod_{j=1}^{J}(q_j)^{b_j} \lim_{r\to\infty}\binom{k_{r,m}}{b_1, b_2, \ldots, b_J}$$

and correspondingly, (38) is changed to

$$\lim_{r\to\infty} \mathbb{E}_{\mathcal{C}}\left[\prod_{j=1}^{J}(g_{j,p})^{\beta_{r,j,m}}\right] \geq \Gamma\left(\sum_{j=1}^{J} q_j g_{j,p}\right).$$

Therefore, the upper bound on the design rate in (32) holds for the more general setting as above.

# 5 Achievable Rates of Punctured LDPC Codes

In this section we derive upper bounds on the achievable rates of punctured LDPC codes whose transmission takes place over an MBIOS channel, and the codes are ML decoded. The analysis in this section relies on the bound presented in Section 4.

Let $\mathcal{C}$ be a binary linear block code. Assume its code bits are partitioned into $J$ disjoint sets, and the bits of the $j^{\text{th}}$ set are randomly punctured with a puncturing rate $\pi_j$ (where $j \in \{1, \ldots, J\}$). The transmission of this code over an MBIOS channel is equivalent to transmitting the code over a set of $J$ parallel MBIOS channels where each of these channels forms a serial concatenation of a BEC whose erasure probability is equal to the puncturing rate $\pi_j$, followed by the original MBIOS channel (see e.g., [4, 10, 11, 12]).

## 5.1 Some Preparatory Lemmas

This sub-section presents two lemmas which are later used to prove results for ensembles of randomly and intentionally punctured LDPC codes (denoted by RP-LDPC and IP-LDPC codes, respectively).

In the following lemma, we consider a punctured linear block code and provide an upper bound on the conditional entropy of a codeword before puncturing, given the received sequence at the output of the channel. This upper bound is expressed in terms of the bit error probability of the punctured code.

**Lemma 5.1.** Let $\mathcal{C}'$ be a binary linear block code of length $n$ and rate $R'$, and let $\mathcal{C}$ be a code which is obtained from $\mathcal{C}'$ by puncturing some of its code bits. Assume that the transmission of the code $\mathcal{C}$ takes place over an arbitrary communication channel, and the code is decoded by an arbitrary decoding algorithm. Let $\mathbf{x}' = (x'_1, \ldots, x'_n)$ and $\mathbf{y} = (y_1, \ldots, y_n)$ designate the transmitted codeword of $\mathcal{C}'$ and the received sequence, respectively. Then, the conditional entropy of the original codeword of $\mathcal{C}'$ given the received sequence satisfies

$$\frac{H(\mathbf{X}'|\mathbf{Y})}{n} \leq R'\, h_2(P_\mathrm{b}) \tag{42}$$

where $P_\mathrm{b}$ designates the bit error probability of the punctured code $\mathcal{C}$.

*Proof.* The proof follows directly from Lemma 3.2, and the equivalence between the transmission of punctured codes over an MBIOS channel and the particular case of transmitting this code over a set of parallel channels (see the introductory paragraph of Section 5). □

Puncturing serves to increase the rate of the original code by reducing the length of the codeword. It may however cause several codewords to be mapped onto a single codeword, thereby reducing the dimension of the code. Consider a binary linear code, $\mathcal{C}'$, of length $n$ and rate $R'$ and assume a fraction $\gamma$ of its code bits are punctured. In the case that the dimension is not reduced by puncturing, the rate of the punctured code is given by $R = \frac{R'}{1-\gamma}$. In the general case, we cannot guarantee that the dimension of the code is not reduced. However, for a sequence of punctured codes whose bit error probability vanishes as the block length of the codes goes to infinity, the following lemma shows that the rate of the punctured codes converges to the desired rate $R$.

**Lemma 5.2.** Let $\{\mathcal{C}'_r\}$ be a sequence of binary linear block codes of length $n_r$ and rate $R'_r$, and let $\{\mathcal{C}_r\}$ be a sequence of codes which is obtained from $\{\mathcal{C}'_r\}$ by puncturing a fraction $\gamma$ of the code bits. Assume the sequence of punctured codes $\{\mathcal{C}_r\}$ achieves vanishing bit error probability in the limit where we let $r$ tend to infinity. Then, the asymptotic rate $R$ of the sequence of punctured codes is given by

$$R = \frac{R'}{1-\gamma} \tag{43}$$

where $R' = \lim_{r\to\infty} R'_r$ is the asymptotic rate of the original sequence of codes $\{\mathcal{C}'_r\}$.

*Proof.* Let $\mathbf{x}'_r = (x'_1, \ldots, x'_{n_r})$ and $\mathbf{y}_r = (y_r \ldots, y_{n_r})$ designate the original codeword (before puncturing) and the received sequence (after puncturing), respectively. Since we assume the there exists a decoding algorithm such that the punctured codes achieve vanishing bit error probability, we have from lemma 5.1 that

$$\lim_{r\to\infty} \frac{H(\mathbf{X}'_r|\mathbf{Y}_r)}{n_r} = 0.$$

Let $\mathbf{x}_r = (x_1, \ldots, x_{n_r})$ designate the codeword after puncturing (where the punctured bits are replaced by question marks). Since $\mathbf{X}'_r \Rightarrow \mathbf{X}_r \Rightarrow \mathbf{Y}_r$ forms a Markov chain, then by the information processing inequality, we get $H(\mathbf{X}'_r|\mathbf{X}_r) \leq H(\mathbf{X}'_r|\mathbf{Y}_r)$. The non-negativity of the conditional entropy therefore yields that

$$\lim_{r\to\infty} \frac{H(\mathbf{X}'_r|\mathbf{X}_r)}{n_r} = 0. \tag{44}$$

Denote the dimension of the codes $\mathcal{C}'_r$ and $\mathcal{C}_r$ by $d'_r$ and $d_r$, respectively. Since $\mathcal{C}'_r$ is binary and linear, every codeword of $\mathcal{C}_r$ originates from exactly $2^{d'_r - d_r}$ different codewords of $\mathcal{C}'_r$. The codewords are assumed to be transmitted with equal probability, and therefore $H(\mathbf{X}'_r|\mathbf{X}_r) = d'_r - d_r$. Let $R_r$ designate the rate of the punctured code $\mathcal{C}_r$. By definition, $d'_r = R'_r n_r$, and since $n_r(1-\gamma)$ forms the block length of the punctured code $\mathcal{C}_r$, then $d_r = R_r n_r (1-\gamma)$. Substituting the last three equalities into (44) gives

$$\lim_{r\to\infty} \big(R'_r - R_r(1-\gamma)\big) = 0.$$

This completes the proof of the lemma. □

For a sequence of codes $\{\mathcal{C}'_r\}$, it is natural to refer to their code rates $R'_r$. However, for sequences of ensembles, where parity-check matrices are randomly picked, such matrices are unlikely to be full rank. Hence, a more natural approach is to refer to their design rates. To this end, we define the design rate of codes which are obtained by puncturing some code bits of binary linear block codes.

**Definition 5.1.** Let $\mathcal{C}'$ be a binary linear block code of length $n$, $H'$ be a $c \times n$ parity-check matrix of $\mathcal{C}'$ and $R'_d \triangleq 1 - \frac{c}{n}$ designate the design rate of the code $\mathcal{C}'$. Let $\mathcal{C}$ be a code which is obtained from $\mathcal{C}'$ by puncturing a fraction $\gamma$ of the code bits. The *design rate* of $\mathcal{C}$ is defined as

$$R_{\mathrm{d}} \triangleq \frac{R'_{\mathrm{d}}}{1-\gamma}. \tag{45}$$

From Lemma 5.2, it follows that for an arbitrary sequence of punctured codes which achieves vanishing bit error probability, their asymptotic design rate is equal in probability 1 to their asymptotic rate if and only if this condition also holds for the original sequence of codes before their puncturing. For un-punctured ensembles of LDPC codes, a sufficient condition for the asymptotic convergence of the rate to the design rate is introduced in [8, Lemma 7] (see Lemma 2.1 in Section 2). In Section 5.4, we apply this lemma to show that the bounds on the achievable rates of ensembles of punctured LDPC codes apply to their actual code rates and not only to their asymptotic design rates.

## 5.2 Randomly Punctured LDPC Codes

In this section, we consider the achievable rates of randomly punctured LDPC (RP-LDPC) codes. We assume that the transmission of these codes takes place over an MBIOS channel, and refer to their achievable rates under optimal ML decoding. The upper bound on the achievable rates of ensembles of RP-LDPC codes relies on the analysis in Section 4 where we derived an upper bound on the achievable rates of LDPC codes for parallel channels.

In the following, we assume that the communication takes place over an MBIOS channel with capacity $C$, and we define

$$g_p \triangleq \int_0^\infty a(l)\, (1+e^{-l}) \tanh^{2p}\left(\frac{l}{2}\right) dl\ , \qquad p \in \mathbb{N} \tag{46}$$

where $a$ designate the *pdf* of the LLR of the channel given that its input is zero.

**Theorem 5.1.** Let $\{(n_r, \lambda, \rho)\}_{r=1}^\infty$ be a sequence of ensembles of LDPC codes whose block length $(n_r)$ tends to infinity as $r \to \infty$. Assume that a sequence of ensembles of RP-LDPC codes is constructed in the following way: for each code from an ensemble of the original sequence, a subset of $\alpha n_r$ code bits is a-priori selected, and these bits are randomly punctured at a fixed rate $(P_{\mathrm{pct}})$. Assume that the punctured codes are transmitted over an MBIOS channel with capacity $C$, and that in the limit where $r$ approaches infinity, the sequence of ensembles of RP-LDPC codes achieves vanishing bit error probability under some decoding algorithm. Then in probability 1 w.r.t. the random puncturing patterns, the asymptotic design rate $(R_{\mathrm{d}})$ of the new sequence satisfies

$$R_{\mathrm{d}} \leq \frac{1}{1-\alpha P_{\mathrm{pct}}} \left( 1 - \frac{1-(1-\alpha P_{\mathrm{pct}})C}{1 - \frac{1}{2\ln 2} \sum_{p=1}^{\infty} \left\{ \frac{1}{p(2p-1)}\, \Gamma\Big( (1-P_{\mathrm{pct}}+\xi) g_p \Big) \right\}} \right) \tag{47}$$

where $\Gamma$, as introduced in (3), denotes the right degree distribution (from the node perspective) of the original sequence, $g_p$ is introduced in (46), and $\xi$ is the following positive number:

$$\xi \triangleq 2(1-\alpha)P_{\text{pct}} \int_0^1 \lambda(x)\, dx. \tag{48}$$

*Proof.* By assumption, we select a set of code bits whose size is a fraction $\alpha$ of the $n_r$ code bits, and these bits are randomly punctured at rate $P_{\text{pct}}$. The transmission of the resulting codeword over an MBIOS channel is equivalent to the transmission of the original codeword over a set of $J = 2$ parallel channels. The first channel, referring to the set of code bits which are randomly punctured, is a serial concatenation of a BEC with erasure probability $P_{\text{pct}}$ and the original MBIOS channel; the second channel which refers to the rest of the bits (which are transmitted without being randomly punctured) is the original MBIOS channel. For simplicity, let us first assume that the degree distribution associated with the selected subset of $\alpha n_r$ code bits which are randomly punctured is independent of the specific code from the ensemble $(n_r, \lambda, \rho)$. Based on the discussion above and the notation in Section 4, the transmission of the $n_r$ code bits over these two parallel channels induces a sequence of ensembles of LDPC codes, $\{(n_r, \underline{\lambda}_r, \rho)\}_{r=1}^{\infty}$, where $\underline{\lambda}_r = (\lambda_r^{[1]}, \lambda_r^{[2]})$ depends on the selection of the subset of $\alpha n_r$ code bits which are randomly punctured. Following this equivalence, we get from the notation in Theorem 4.1 that

$$\begin{aligned} & p_1 = \alpha, \quad p_2 = 1-\alpha, \quad C_1 = C(1-P_{\text{pct}}), \quad C_2 = C \\ \Rightarrow \quad & \sum_{j=1}^{J} p_j C_j = C(1-\alpha P_{\text{pct}}). \end{aligned} \tag{49}$$

In order to apply Theorem 4.1 to our case, we find a global lower bound on the sum $\sum_{j=1}^{J} q_j g_{j,p}$ which does not depend on the a-priori selection of the subset of code bits which are randomly punctured. From (8) and (46), it follows that for all $p \in \mathbb{N}$:

$$\begin{aligned} g_{1,p} &= \int_0^\infty \left[P_{\text{pct}}\delta(l) + (1-P_{\text{pct}})a(l)\right](1+e^{-l}) \tanh^{2p}\left(\frac{l}{2}\right) dl \\ &= (1-P_{\text{pct}}) \int_0^\infty a(l)(1+e^{-l}) \tanh^{2p}\left(\frac{l}{2}\right) dl \\ &= (1-P_{\text{pct}})\, g_p \end{aligned}$$

and $g_{2,p} = g_p$. Based on Lemmas 4.1 and 4.2, we get that for all $p \in \mathbb{N}$

$$q_1 g_{1,p} + q_2 g_{2,p} = \frac{\alpha g_p (1-P_{\text{pct}}) \int_0^1 \lambda(x) dx}{\int_0^1 \lambda_r^{[1]}(x) dx} + \frac{(1-\alpha) g_p \int_0^1 \lambda(x) dx}{\int_0^1 \lambda_r^{[2]}(x) dx} \tag{50}$$

where the following constraint is satisfied:

$$\frac{\alpha}{\int_0^1 \lambda_r^{[1]}(x) dx} + \frac{1-\alpha}{\int_0^1 \lambda_r^{[2]}(x) dx} = \frac{1}{\int_0^1 \lambda(x) dx} \tag{51}$$

and

$$\int_0^1 \lambda_r^{[1]}(x)\, dx \leq \frac{1}{2}\,, \quad \int_0^1 \lambda_r^{[2]}(x)\, dx \leq \frac{1}{2} \tag{52}$$

due to the fact that $\lambda^{[1]}(x) \le x$ and $\lambda^{[2]}(x) \le x$ for $x \in [0,1]$ (even without explicitly knowing $\lambda^{[1]}$ and $\lambda^{[2]}$ which depend on the a-priori choice of the subset of bits which are randomly punctured). Based on (50)–(52), we get

$$\begin{aligned}
& q_1 g_{1,p} + q_2 g_{2,p} \\
&= (1-P_{\text{pct}}) g_p \int_0^1 \lambda(x)\, dx \left( \frac{\alpha}{\int_0^1 \lambda_r^{[1]}(x) dx} + \frac{1-\alpha}{\int_0^1 \lambda_r^{[2]}(x) dx} \right) + \frac{(1-\alpha) P_{\text{pct}} g_p \int_0^1 \lambda(x)\, dx}{\int_0^1 \lambda^{[2]}(x) dx} \\
&= (1-P_{\text{pct}}) g_p + \frac{(1-\alpha) P_{\text{pct}} g_p \int_0^1 \lambda(x)\, dx}{\int_0^1 \lambda_r^{[2]}(x) dx} \\
&\ge \left(1 - P_{\text{pct}} + 2(1-\alpha) P_{\text{pct}} \int_0^1 \lambda(x)\, dx \right) g_p \\
&= (1 - P_{\text{pct}} + \xi) g_p
\end{aligned}$$

where $\xi$ is defined in (48). Since the degree distribution $\Gamma$ is a monotonic increasing function, then

$$\Gamma\left(\sum_{j=1}^{J} q_j g_{j,p}\right) \ge \Gamma\big((1-P_{\text{pct}}+\xi) g_p\big). \tag{53}$$

By substituting (49) and (53) in the RHS of (32), we obtain the following upper bound on the asymptotic design rate of the original sequence

$$R'_{\text{d}} \le 1 - \frac{1-(1-\alpha P_{\text{pct}})C}{1 - \frac{1}{2\ln 2} \sum_{p=1}^{\infty} \left\{ \frac{1}{p(2p-1)}\, \Gamma\Big((1-P_{\text{pct}}+\xi) g_p\Big) \right\}}.$$

Since as $r \to \infty$, in probability 1 w.r.t. the puncturing patterns, a fraction $\gamma = \alpha P_{\text{pct}}$ of the code bits are punctured, then the asymptotic design rate ($R_{\text{d}}$) of this sequence satisfies the equality

$$R_{\text{d}} = \frac{R'_{\text{d}}}{1-\alpha P_{\text{pct}}} \tag{54}$$

from which the theorem follows.

For the case where the degree distribution associated with the subset of code bits which are randomly punctured depends on the code $\mathcal{C}$ from the ensemble $(n_r, \lambda, \rho)$, the pair $(\lambda_r^{[1]}, \lambda_r^{[2]})$ cannot be considered to be uniform over all the codes from this ensemble. In this case, Theorem 4.1 is not directly applicable. In order to circumvent the problem, we rely on the discussion in Remark 4.1, and on the fact that the lower bound on $q_1 g_{1,p} + q_2 g_{2,p}$ which is given above in terms of $\xi$ from (48) is universal for all the codes from this ensemble (i.e., it only depends on $\lambda$, but does not depend on the specific degree distributions $\lambda_r^{[1]}(\mathcal{C})$ and $\lambda_r^{[2]}(\mathcal{C})$ which are associated with the code $\mathcal{C}$ from the ensemble). In light of this reasoning, the proof of the theorem for ensembles of RP-LDPC codes also follows in the more general setting where the degree distribution associated with the subset of the code bits which are randomly punctured depends on the specific code from the ensemble. □

### 5.3 Intentionally Punctured LDPC Codes

In [4], Ha and McLaughlin show that good codes can be constructed by puncturing good ensembles of LDPC codes using a technique called "intentional puncturing". In this approach, the code bits are partitioned into disjoint sets so that each set contains all the code bits whose corresponding variable nodes have the same degree. The code bits in each one of these sets are randomly punctured at a fixed puncturing rate.

We briefly present the notation used in [4] for the characterization of ensembles of intentionally punctured LDPC (IP-LDPC) codes. Consider an ensemble of LDPC codes with left and right edge degree distributions $\lambda$ and $\rho$, respectively. For each degree $j$ such that $\lambda_j > 0$, a puncturing rate $\pi_j \in [0, 1]$ is determined for randomly puncturing the set of code bits which correspond to variable nodes of degree $j$. The polynomial associated with this puncturing pattern is

$$\pi^0(x) \triangleq \sum_{j=1}^{\infty} \pi_j x^{j-1}. \tag{55}$$

An ensemble of IP-LDPC codes can be therefore represented by the quadruplet $(n, \lambda, \rho, \pi^0)$ where $n$ designates the block length of these codes, $\lambda$ and $\rho$ are the left and right degree distributions from the edge perspective, respectively, and $\pi^{(0)}$ is the polynomial which corresponds to the puncturing pattern, as given in (55). The average fraction of punctured bits is given by $p^{(0)} = \sum_{j=1}^{\infty} \Lambda_j \pi_j$ where $\Lambda$ is the left node degree distribution of the original LDPC ensemble. The following statement, which relies on Theorem 4.1, provides an upper bound on the common design rate of a sequence of ensembles of IP-LDPC codes. This bound refers to ML decoding (and hence, to any sub-optimal decoding algorithm).

**Theorem 5.2.** Let $\{(n_r, \lambda, \rho, \pi^0)\}_{r=1}^{\infty}$ be a sequence of ensembles of IP-LDPC codes transmitted over an MBIOS channel, and assume that $n_r$ tends to infinity as $r \to \infty$. Let $C$ be the channel capacity, and $a$ be the *pdf* of the LLR at the output of the channel given its input is 1. If the asymptotic bit error probability of this sequence vanishes under ML decoding (or any sub-optimal decoding algorithm) as $r \to \infty$, then in probability 1 w.r.t. the puncturing patterns, the common design rate $R_{\text{d}}$ of these ensembles satisfies

$$R_{\text{d}} \leq \frac{1}{1 - p^{(0)}} \left[ 1 - \frac{1 - (1 - p^{(0)})C}{1 - \frac{1}{2 \ln 2} \sum_{p=1}^{\infty} \left\{ \frac{1}{p(2p-1)} \, \Gamma\left( \left(1 - \sum_{j=1}^{\infty} \lambda_j \pi_j \right) g_p \right) \right\}} \right] \tag{56}$$

where $\Gamma$, as introduced in (3), denotes the right degree distribution from the node perspective,

$$p^{(0)} \triangleq \sum_{j=1}^{\infty} \Lambda_j \pi_j \tag{57}$$

designates the average puncturing rate of the code bits, and $g_p$ is the functional of the MBIOS channel introduced in (46).

*Proof.* The proof follows from Theorem 4.1, and the observation that IP-LDPC codes form a special case of the ensemble $(n, \underline{\lambda}, \rho)$ examined in Section 4. For a sequence of ensembles of IP-LDPC codes, $\{(n_r, \lambda, \rho, \pi^0)\}$, the number of parallel MBIOS channels used for transmission is equal to the number of strictly positive coefficients in the polynomial $\lambda$, i.e., $J \triangleq |\{i : \lambda_i > 0\}|$. Denote these degrees by $i_1, \ldots, i_J$, then the bits transmitted over the $j^{\text{th}}$ channel are those involved in exactly $i_j$ parity-check equations (i.e., the bits whose corresponding variable nodes are of degree $i_j$). From the above discussion, it follows that the fraction of code bits transmitted over the $j^{\text{th}}$ channel is given by

$$p_j = \Lambda_{i_j} \,, \quad j \in \{1, \ldots, J\} \tag{58}$$

and the fraction of edges in the bipartite graph which are connected to variable nodes transmitted of the $j^{\text{th}}$ channel is given by

$$q_j = \lambda_{i_j}\ , \quad j \in \{1, \ldots, J\}. \tag{59}$$

The transmission of IP-LDPC codes over an MBIOS channel is equivalent to transmitting these code over a set of $J$ parallel MBIOS channels where each of these channels forms a serial concatenation of a BEC whose erasure probability is equal to the puncturing rate $\pi_{i_j}$, followed by the original MBIOS channel. Hence, the *pdf* of the LLR at the output of the $j^{\text{th}}$ MBIOS channel given its input is 1 gets the form

$$a(l; j) = \pi_{i_j} \delta_0(l) + (1 - \pi_{i_j}) a(l), \quad l \in \mathbb{R} \tag{60}$$

and the capacity of this channel is

$$C_j = C(1 - \pi_{i_j}). \tag{61}$$

By substituting (60) into (8), we get that for all $j \in \{1, \ldots, J\}$ and $p \in \mathbb{N}$

$$\begin{aligned} g_{j,p} &= \int_0^\infty \left[\pi_{i_j}\delta_0(l) + (1-\pi_{i_j})a(l)\right]\ (1+e^{-l}) \tanh^{2p}\left(\frac{l}{2}\right)\ \mathrm{d}l \\ &= (1-\pi_{i_j}) \int_0^\infty a(l)\ (1+e^{-l}) \tanh^{2p}\left(\frac{l}{2}\right)\ \mathrm{d}l \\ &= (1-\pi_{i_j}) g_p \end{aligned} \tag{62}$$

where the last equality is based on (46). The statement now follows by substituting (58), (59), (61) and (62) in (32); we finally use the scaling factor for the design rate of punctured codes, as given in Definition 5.1 (in this case, the parameter $\gamma$ introduced in this definition tends to $p^{(0)}$ where this convergence is in probability 1 w.r.t. the puncturing patterns; $p^{(0)}$, as introduced in (57), denotes the average puncturing rate of the code bits). Finally, since $\lambda_j = \Lambda_j = 0$ for $j \notin \{i_1, \ldots, i_J\}$, then regarding the sums in the RHS of (56), we get the equalities $\sum_{j=1}^{\infty} \Lambda_j \pi_j = \sum_{j=1}^{J} \Lambda_{i_j} \pi_{i_j}$ and $\sum_{j=1}^{\infty} \lambda_j \pi_j = \sum_{j=1}^{J} \lambda_{i_j} \pi_{i_j}$. This completes the proof of the theorem. □

## 5.4 Numerical Results for Intentionally Punctured LDPC Codes

In this section, we present a comparison between thresholds under message-passing iterative (MPI) decoding and bounds on thresholds under ML decoding for ensembles of IP-LDPC codes. It is assumed that the transmission of the punctured LDPC codes takes place over a binary-input AWGN channel. The pairs of degree distributions and the corresponding puncturing patterns were originally presented in [4, 5]. We use these ensembles in order to study their inherent gap to capacity, and also study how close to optimal is iterative decoding for these ensembles (in the asymptotic case where the block length goes to infinity).

We refer here to three ensembles of IP-LDPC codes: Tables 1 and 2 refer to two ensembles of rate$-\frac{1}{2}$ LDPC codes which by puncturing, their rates vary between 0.50 and 0.91; Table 3 refers to an ensemble of rate$-\frac{1}{10}$ LDPC codes which by puncturing, its rate varies between 0.10 and 0.83. Based on Lemma 2.1, we verify that the design rates of these three ensembles of LDPC codes (before puncturing) are equal in probability 1 to the asymptotic rates of codes from these ensembles. This conclusion still holds for the punctured LDPC ensembles given in Tables 1–3 (see Lemma 5.2). This enables to calculate the capacity limits which refer to the design rates of these ensembles, and to evaluate the gaps to capacity under ML decoding and iterative decoding for these ensembles of punctured LDPC codes.

For various ensembles of IP-LDPC codes, Tables 1–3 provide lower bounds on the inherent gap to capacity under optimal ML decoding (based on Theorem 5.2); these values are compared to the corresponding gaps to capacity under iterative message-passing decoding (whose calculation is based on the density evolution analysis). On one hand, Tables 1–3 provide a quantitative assessment of the loss in the asymptotic performance which is attributed to the sub-optimality of iterative decoding (as compared to optimal ML decoding), and on the other hand, they provide an assessment of the inherent loss in performance which is attributed to the structure of the ensembles, even if optimal ML decoding could be applied to decode these codes. The loss in performance in both cases is measured in terms of $\frac{E_b}{N_0}$ in decibels. It is demonstrated in Tables 1–3 that for various good ensembles of IP-LDPC codes, the asymptotic loss in performance due to the code structure is still non-negligible as compared to the corresponding loss due to the sub-optimality of iterative decoding. As an example, for all the ensembles of IP-LDPC codes considered in Table 1 (which were originally introduced in [4, Table 2]), the gap to capacity under the sum-product iterative decoding algorithm does not exceed 0.6 dB; however, under ML decoding, the gap to capacity is always greater than $\frac{1}{3}$ of the corresponding gap to capacity under this iterative decoding algorithm; therefore, the results in Table 1 regarding the thresholds under ML decoding further emphasize the efficiency of the sum-product decoding algorithm for these ensembles, especially in light of its moderate complexity.

| $\pi^0(x)$ (puncturing pattern) | Design rate | Capacity limit | Lower bound (ML decoding) | Iterative (IT) Decoding | Fractional gap to capacity (ML vs. IT) |
|---|---|---|---|---|---|
| 0 | 0.500 | 0.187 dB | 0.270 dB | 0.393 dB | $\geq 40.3\%$ |
| $0.07886x + 0.01405x^2 + 0.06081x^3 + 0.07206x^9$ | 0.528 | 0.318 dB | 0.397 dB | 0.526 dB | $\geq 37.9\%$ |
| $0.20276x + 0.09305x^2 + 0.03356x^3 + 0.16504x^9$ | 0.592 | 0.635 dB | 0.716 dB | 0.857 dB | $\geq 36.4\%$ |
| $0.25381x + 0.15000x^2 + 0.34406x^3 + 0.019149x^9$ | 0.629 | 0.836 dB | 0.923 dB | 1.068 dB | $\geq 37.3\%$ |
| $0.31767x + 0.18079x^2 + 0.05265x^3 + 0.24692x^9$ | 0.671 | 1.083 dB | 1.171 dB | 1.330 dB | $\geq 35.6\%$ |
| $0.36624x + 0.24119^2 + 0.49649x^3 + 0.27318x^9$ | 0.719 | 1.398 dB | 1.496 dB | 1.664 dB | $\geq 36.9\%$ |
| $0.41838x + 0.29462x^2 + 0.05265x^3 + 0.30975x^9$ | 0.774 | 1.814 dB | 1.927 dB | 2.115 dB | $\geq 37.2\%$ |
| $0.47074x + 0.34447x^2 + 0.02227x^3 + 0.34997x^9$ | 0.838 | 2.409 dB | 2.547 dB | 2.781 dB | $\geq 37.1\%$ |
| $0.52325x + 0.39074x^2 + 0.01324x^3 + 0.39436x^9$ | 0.912 | 3.399 dB | 3.607 dB | 3.992 dB | $\geq 35.1\%$ |

Table 1: Comparison of thresholds for ensembles of IP-LDPC codes where the original ensemble before puncturing has the degree distributions $\lambda(x) = 0.25105x + 0.30938x^2 + 0.00104x^3 + 0.43853x^9$ and $\rho(x) = 0.63676x^6 + 0.36324x^7$ (so its design rate is equal to $\frac{1}{2}$). The transmission of these codes takes place over a binary-input AWGN channel. The table compares values of $\frac{E_b}{N_0}$ referring to the capacity limit, the bound given in Theorem 5.2 (which provides a lower bound on $\frac{E_b}{N_0}$ under ML decoding), and thresholds under iterative message-passing decoding. The fractional gap to capacity (see the rightmost column) measures the ratio of the gap to capacity under optimal ML decoding and the achievable gap to capacity under (sub-optimal) iterative message-passing decoding. The pair of degree distributions for the ensemble of LDPC codes, and the polynomials which correspond to its puncturing patterns are taken from [4, Table 2].

Tables 1–3 also show that the performance of the punctured LDPC codes is degraded at high rates, where one needs to pay a considerable penalty for using punctured codes. This phenomenon was explained in [12, Theorem 1] by the threshold effect for ensembles of IP-LDPC codes.

Following the performance analysis of punctured LDPC codes in [2, 4, 5, 12], the numerical results shown in Tables 1–3 exemplify the high potential of puncturing in designing codes which operate closely to the Shannon capacity limit and used for rate-compatible coding for various MBIOS channels. Other examples of capacity-achieving ensembles of punctured codes on graphs are the irregular repeat-accumulate (IRA) codes and accumulate-repeat-accumulate (ARA) codes. Recently, it was shown by Pfister et al. that properly designed nonsystematic IRA codes achieve the capacity of the BEC with bounded decoding complexity per information bit [10]. This bounded complexity result is achieved by puncturing all the information bits of the IRA codes, and allowing in this way a sufficient number of state nodes in the Tanner graph representing the codes. This is in contrast to all previous constructions of capacity-achieving LDPC codes whose complexity becomes unbounded as their gap to capacity vanishes.

| $\pi^0(x)$ (puncturing pattern) | Design rate | Capacity limit | Lower bound (ML decoding) | Iterative (IT) Decoding | Fractional gap to capacity (ML vs. IT) |
|---|---|---|---|---|---|
| 0 | 0.500 | 0.187 dB | 0.234 dB | 0.299 dB | $\geq 41.5\%$ |
| $0.102040x + 0.06497x^2 + 0.06549x^5 + 0.00331x^6 + 0.39377x^{19}$ | 0.555 | 0.450 dB | 0.473 dB | 0.599 dB | $\geq 15.4\%$ |
| $0.226410x + 0.14149x^2 + 0.21268x^5 + 0.00001x^6 + 0.4424x^{19}$ | 0.625 | 0.816 dB | 0.841 dB | 1.028 dB | $\geq 11.9\%$ |
| $0.348940x + 0.21015x^2 + 0.38902x^5 + 0.00003x^6 + 0.48847x^{19}$ | 0.714 | 1.368 dB | 1.398 dB | 1.699 dB | $\geq 8.9\%$ |
| $0.410320x + 0.24330x^2 + 0.48388x^5 + 0.00004x^6 + 0.50541x^{19}$ | 0.769 | 1.777 dB | 1.811 dB | 2.215 dB | $\geq 7.8\%$ |
| $0.469100x + 0.28408x^2 + 0.56178x^5 + 0.00002x^6 + 0.53412x^{19}$ | 0.833 | 2.362 dB | 2.404 dB | 3.004 dB | $\geq 6.6\%$ |
| $0.533750x + 0.30992x^2 + 0.66375x^5 + 0.00001x^6 + 0.54837x^{19}$ | 0.909 | 3.343 dB | 3.410 dB | 4.634 dB | $\geq 5.2\%$ |

Table 2: Comparison of thresholds for ensembles of IP-LDPC codes where the original LDPC ensemble before puncturing has the degree distributions $\lambda(x) = 0.23403x + 0.21242x^2 + 0.14690x^5 + 0.10284x^6 + 0.30381x^{19}$ and $\rho(x) = 0.71875x^7 + 0.28125x^8$ (so its design rate is equal to $\frac{1}{2}$). The transmission of these codes takes place over a binary-input AWGN channel. The table compares values of $\frac{E_b}{N_0}$ referring to the capacity limit, the bound given in Theorem 5.2 (which provides a lower bound on $\frac{E_b}{N_0}$ under ML decoding), and thresholds under iterative message-passing decoding. The fractional gap to capacity (see the rightmost column) measures the ratio of the gap to capacity under optimal ML decoding and the achievable gap to capacity under (sub-optimal) iterative message-passing decoding. The pair of degree distributions for the ensemble of LDPC codes, and the polynomials which correspond to the puncturing patterns are taken from [4, Table 3].

The decoding complexity of punctured LDPC codes for parallel channels is addressed in the next section.

| $\pi^0(x)$ (puncturing pattern) | Design rate | Capacity limit | Lower bound (ML decoding) | Iterative (IT) Decoding | Fractional gap to capacity (ML vs. IT) |
|---|---|---|---|---|---|
| 0 | 0.100 | $-1.286$ dB | $-1.248$ dB | $-1.028$ dB | $\geq 14.5\%$ |
| $0.486490x + 0.69715x^2 + 0.03287x^3 + 0.04248x^4 + 0.69048x^7 + 0.45209x^{24}$ | 0.203 | $-0.953$ dB | $-0.917$ dB | $-0.731$ dB | $\geq 16.3\%$ |
| $0.655580x + 0.83201x^2 + 0.48916x^3 + 0.33917x^4 + 0.63990x^7 + 0.76947x^{24}$ | 0.304 | $-0.605$ dB | $-0.570$ dB | $-0.317$ dB | $\geq 12.0\%$ |
| $0.745690x + 0.87184x^2 + 0.38179x^3 + 0.48427x^4 + 0.74655x^7 + 0.79130x^{24}$ | 0.406 | $-0.226$ dB | $-0.189$ dB | $+0.029$ dB | $\geq 14.7\%$ |
| $0.838470x + 0.65105x^2 + 0.04527x^3 + 0.95233x^4 + 0.74808x^7 + 0.80845x^{24}$ | 0.487 | $+0.130$ dB | $+0.171$ dB | $+0.599$ dB | $\geq 8.7\%$ |
| $0.979320x + 0.46819x^2 + 0.71050x^3 + 0.59816x^4 + 0.79485x^7 + 0.05765x^{24}$ | 0.577 | $+0.556$ dB | $+0.840$ dB | $+1.152$ dB | $\geq 47.7\%$ |
| $0.895200x + 0.84401x^2 + 0.98541x^3 + 0.42518x^4 + 0.92976x^7 + 0.30225x^{24}$ | 0.663 | $+1.039$ dB | $+1.232$ dB | $+1.806$ dB | $\geq 25.2\%$ |
| $0.910960x + 0.91573x^2 + 0.23288x^3 + 0.40977x^4 + 0.99811x^7 + 0.15915x^{24}$ | 0.747 | $+1.605$ dB | $+1.958$ dB | $+2.637$ dB | $\geq 34.2\%$ |
| $0.904130x + 0.96192x^2 + 0.35996x^3 + 0.96980x^4 + 0.31757x^7 + 0.89250x^{24}$ | 0.828 | $+2.303$ dB | $+2.505$ dB | $+3.863$ dB | $\geq 13.0\%$ |

Table 3: Comparison of thresholds for ensembles of IP-LDPC codes where the original ensemble before puncturing has the degree distributions $\lambda(x) = 0.414936x + 0.183492x^2 + 0.013002x^3 + 0.093081x^4 + 0.147017x^7 + 0.148472x^{24}$ and $\rho(x) = 0.4x^2 + 0.6x^3$ (so its design rate is equal to $\frac{1}{10}$). The transmission of these codes takes place over a binary-input AWGN channel. The table compares values of $\frac{E_b}{N_0}$ referring to the capacity limit, the bound given in Theorem 5.2 (which provides a lower bound on $\frac{E_b}{N_0}$ under ML decoding), and thresholds under iterative message-passing decoding. The fractional gap to capacity (see the rightmost column) measures the ratio of the gap to capacity under optimal ML decoding and the achievable gap to capacity under (sub-optimal) iterative message-passing decoding. The pair of degree distributions for the ensemble of LDPC codes, and the polynomials which correspond to the puncturing patterns are taken from [5, Table 5.1].

# 6 Lower Bounds on the Decoding Complexity of LDPC Codes for Parallel Channels

The scope of this section is to derive a lower bound on the decoding complexity of LDPC codes for parallel MBIOS channels. The lower bound holds under message-passing iterative (MPI) decoding, and it grows like the logarithm of the inverse of the gap (in rate) to capacity. Interestingly, a logarithmic behavior of the parity-check density (which forms a measure of the decoding complexity per iteration) in terms of the gap to capacity also characterizes the upper bound derived in [3, Section 3]; this upper bound refers to MacKay's ensemble of LDPC codes whose transmission takes place over a set of parallel MBIOS channels.

In the previous section we regarded the transmission of punctured LDPC codes over MBIOS channels as a particular case of the transmission of the original codes (before puncturing) over a set of parallel MBIOS channels. Hence, the aforementioned bound is later applied to obtain lower bounds on the decoding complexity of (randomly and intentionally) punctured LDPC codes. This section refers to an appendix which suggests a simplified re-derivation of [10, Theorems 3 and 4], and shows that the bounds introduced in this section are tighter.

## 6.1 A Lower Bound on the Decoding Complexity for Parallel MBIOS Channels

Consider a binary linear block code which is represented by a bipartite graph, and assume that the graph serves for the decoding with an iterative algorithm. Following [3] and [10], the decoding complexity under MPI decoding is defined as the number of edges in the graph normalized per information bit. This quantity measures the number of messages which are delivered through the edges of the graph (from left to right and vice versa) during a single iteration. Equivalently, since there is a one-to-one correspondence between a bipartite graph and the parity-check matrix $H$ which represents the code, the decoding complexity is also equal to the number of non-zero elements in $H$ normalized per information bit (i.e., the density of the parity-check matrix [15, Definition 2.2]). Hence, the decoding complexity (as well as the performance) of iteratively decoded binary linear block codes depends on the specific representation of the code by a parity-check matrix. Since the average right degree ($a_{\text{R}}$) of a bipartite graph is equal to the number of edges normalized per parity-check equation, then the average right degree and the decoding complexity are related quantities. Consider an ensemble of LDPC codes whose design rate is $R_{\text{d}}$. It is natural to relate the decoding complexity of the ensemble, say $\chi_D$, to its average right degree and design rate, as follows:

$$\chi_{\text{D}} = \frac{1 - R_{\text{d}}}{R_{\text{d}}}\, a_{\text{R}}\ .$$

We note that $a_{\text{R}}$ is fixed for all the codes from an ensemble of LDPC codes with a given pair of degree distributions.

The following lemma will be used for the derivation of a lower bound on the decoding complexity per iteration under MPI decoding.

**Lemma 6.1.** Let $\Gamma$ be the right degree distribution of an ensemble of LDPC codes. Then for any $\alpha \geq 0$ it holds that

$$\Gamma(\alpha) \geq \alpha^{a_{\text{R}}}.$$

*Proof.* Using the convexity of the function $f(x) = \alpha^x$, it follows from Jensen's inequality that

$$\Gamma(\alpha) = \sum_{i=1}^{\infty} \Gamma_i \alpha^i \geq \alpha^{\sum_{i=1}^{\infty} i\Gamma_i} = \alpha^{a_{\text{R}}}.$$

□

Consider a sequence of ensembles of LDPC codes, $\{n_r, \underline{\lambda}, \rho\}_{r=1}^{\infty}$, whose transmission takes place over a set of $J$ statistically independent parallel MBIOS channels. Let $C_j$ and $p_j$ be the capacity and the fraction of code bits which are assigned to the $j^{\text{th}}$ channel, respectively (where $j \in \{1, \ldots, J\}$). We define the average capacity of the set of $J$ parallel channels as $\overline{C} \triangleq \sum_{j=1}^{J} p_j C_j$. For an ensemble of LDPC codes which achieves vanishing bit error probability as the block length goes to infinity, the multiplicative gap (in rate) to capacity is defined as

$$\varepsilon \triangleq 1 - \frac{R_{\text{d}}}{\overline{C}}. \tag{63}$$

We now present a lower bound on the decoding complexity per iteration under MPI decoding for this sequence. The bound is given in terms of the gap to capacity.

**Theorem 6.1.** Let a sequence of ensembles of LDPC codes, $\{(n_r, \underline{\lambda}_r, \rho)\}_{r=1}^{\infty}$, be transmitted over a set of $J$ statistically independent parallel MBIOS channels. Assume that the capacities $C_j$ of these channels are all positive, and denote the average capacity by $\overline{C} \triangleq \sum_{j=1}^{J} p_j C_j$. If this sequence achieves a fraction $1-\varepsilon$ of $\overline{C}$ with vanishing bit error probability, then the asymptotic decoding complexity under MPI decoding satisfies

$$\chi_{\mathrm{D}}(\varepsilon) \geq K_1 + K_2 \ln\left(\frac{1}{\varepsilon}\right) \tag{64}$$

where

$$K_1 = -\frac{(1-\overline{C})\ \ln\left(\frac{1}{2\ln 2}\frac{1-\overline{C}}{\overline{C}}\right)}{\overline{C}\ \ln\left(\sum_{j=1}^{J} q_j g_{j,1}\right)}, \quad K_2 = -\frac{1-\overline{C}}{\overline{C}\ \ln\left(\sum_{j=1}^{J} q_j g_{j,1}\right)} \tag{65}$$

where $g_{j,1}$ is introduced in (8), and $q_j$ is introduced in (31) and is assumed to be positive for all $j \in \{1, \ldots, J\}$. For parallel BECs, the term $\frac{1}{2\ln 2}$ can be removed from the numerator of $K_1$.

*Proof.* Substituting (63) in (32) yields

$$(1-\varepsilon)\overline{C} \leq 1 - \frac{1-\overline{C}}{1 - \frac{1}{2\ln 2}\sum_{p=1}^{\infty}\left\{\frac{1}{p(2p-1)}\,\Gamma\left(\sum_{j=1}^{J} q_j g_{j,p}\right)\right\}}. \tag{66}$$

Since $g_{j,p}$ in (8) is non-negative for $j \in \{1, \ldots, J\}$ and $p \in \mathbb{N}$, and the function $\Gamma$ is non-negative on $\mathbb{R}^+$, then the terms in the infinite sum above are all non-negative. By the truncation of this series where we only take its first term (note that this is the largest term in the sum), we obtain a lower bound on the RHS of (66). This implies that

$$(1-\varepsilon)\overline{C} \leq 1 - \frac{1-\overline{C}}{1 - \frac{1}{2\ln 2}\,\Gamma\left(\sum_{j=1}^{J} q_j g_{j,1}\right)}.$$

Invoking Lemma 6.1 yields that

$$(1-\varepsilon)\overline{C} \leq 1 - \frac{1-\overline{C}}{1 - \frac{1}{2\ln 2}\left(\sum_{j=1}^{J} q_j g_{j,1}\right)^{a_{\mathrm{R}}}}.$$

The solution of the last inequality for the average right degree ($a_{\mathrm{R}}$) gives

$$\begin{aligned} a_{\mathrm{R}} \ &\geq\ -\frac{\ln\left(\frac{1}{2\ln 2}\left(1 + \frac{1-\overline{C}}{\overline{C}\varepsilon}\right)\right)}{\ln\left(\sum_{j=1}^{J} q_j g_{j,1}\right)} \\ &>\ K_1' + K_2' \ln\left(\frac{1}{\varepsilon}\right) \end{aligned} \tag{67}$$

where the last step follows by dropping the 1 which appeared inside the logarithm at the numerator (this step is valid since the denominator is strictly negative), and

$$K_1' = -\frac{\ln\left(\frac{1}{2\ln 2}\frac{1-\overline{C}}{\overline{C}}\right)}{\ln\left(\sum_{j=1}^{J} q_j g_{j,1}\right)}, \quad K_2' = -\frac{1}{\ln\left(\sum_{j=1}^{J} q_j g_{j,1}\right)}\,.$$

Since $R_\mathrm{d} < \overline{C}$, it follows that $\chi_\mathrm{D} = \frac{1-R_\mathrm{d}}{R_\mathrm{d}}\, a_\mathrm{R} > \frac{1-\overline{C}}{\overline{C}}\, a_\mathrm{R}$. The proof of the lower bound on the decoding complexity for parallel MBIOS channels follows by multiplying both sides of (67) by $\frac{1-\overline{C}}{\overline{C}}$.

For parallel BECs, we get from (8) that for every $p \in \mathbb{N}$

$$g_{j,p} = \int_0^\infty a(l;j)(1+e^{-l}) \tanh^{2p}\left(\frac{l}{2}\right) dl = 1 - \varepsilon_j$$

where $\varepsilon_j$ denotes the erasure probability of the $j^{\text{th}}$ BEC. This gives

$$\begin{aligned} &\frac{1}{2\ln 2} \sum_{p=1}^{\infty} \left\{ \frac{1}{p(2p-1)}\ \Gamma\left(\sum_{j=1}^{J} q_j g_{j,p}\right) \right\} \\ &= \frac{1}{2\ln 2} \sum_{p=1}^{\infty} \frac{1}{p(2p-1)} \cdot \Gamma\left(\sum_{j=1}^{J} q_j g_{j,1}\right) \\ &= \Gamma\left(\sum_{j=1}^{J} q_j g_{j,1}\right). \end{aligned}$$

Substituting this in (66), gives

$$(1-\varepsilon)\overline{C} \leq 1 - \frac{1-\overline{C}}{1 - \Gamma\left(\sum_{j=1}^{J} q_j g_{j,1}\right)}.$$

The continuation of the proof follows the same steps as the proof for parallel MBIOS channels, and leads to the improved coefficient $K_1$, i.e., without the factor $\frac{1}{2\ln 2}$ in the numerator of $K_1$ for general MBIOS channels (see (65)). □

We proceed the analysis by the derivation of lower bounds on the decoding complexity of sequences of ensembles of punctured LDPC codes where it is assumed that these sequences achieve vanishing bit error probability; similarly to Theorem 6.1, the lower bounds are expressed in terms of the multiplicative gap (in rate) to capacity.

## 6.2 Lower Bounds on the Decoding Complexity for Punctured LDPC Codes

As discussed in the previous section, transmission of punctured codes can be interpreted as a special case of transmitting the original (un-punctured) codes over a set of parallel channels where these component channels are formed by a mixture of the communication channel and BECs whose erasure probabilities are the puncturing rates of different subsets of code bits. Hence, the bounds on the decoding complexity of punctured codes can be derived as special cases of the bound given in Theorem 6.1. For the sake of brevity, we derive these bounds by using the upper bounds on the achievable rates of punctured LDPC codes given in Theorem 5.1 (for random puncturing) and Theorem 5.2 (for intentional puncturing). Note that the derivation of the latter two theorems relies on Theorem 4.1 (as shown in Fig. 1 on p. 32).

Consider an ensemble of LDPC codes of length $n$ and design rate $R'_\mathrm{d}$, and let the code bits be partitioned into $J$ disjoint sets where the $j^{\text{th}}$ set contains a fraction $p_j$ of these bits ($j \in \{1, \ldots, J\}$). Assume that the bits in the $j^{\text{th}}$ set are randomly punctured at rate $\pi_j$, and let the punctured codes be transmitted over an MBIOS channel whose capacity is $C$. As shown in the previous section, this is equivalent to transmitting the original (un-punctured) codes over a set of $J$ parallel channels, where the $j^{\text{th}}$ set of code bits is transmitted over a channel whose capacity is $C_j = (1-\pi_j)C$. The average capacity of this set of $J$ parallel channels is therefore given by

$$\overline{C} = \sum_{j=1}^{J} p_j \, (1-\pi_j)C = \big(1 - \sum_{j=1}^{J} p_j \pi_j\big)C = (1-\gamma)C \tag{68}$$

where $\gamma \triangleq \sum_{j=1}^{J} p_j \pi_j$ is the overall puncturing rate. Denote the design rate of the punctured codes by $R_{\text{d}} \triangleq \frac{R'_{\text{d}}}{1-\gamma}$ (see Definition 5.1 on p. 18), then it follows that the multiplicative gap to capacity of the punctured codes is given by

$$\varepsilon = 1 - \frac{R_{\text{d}}}{C} = 1 - \frac{R'_{\text{d}}}{\overline{C}}. \tag{69}$$

For punctured codes, the iterative decoder is based on the bipartite graph of the 'mother code' where the channel input to the variable nodes which correspond to the punctured code bits is defined to be 0. Hence, the decoding complexity of the punctured ensemble under MPI decoding is identical to the decoding complexity of the original ensemble (before puncturing), and is given by

$$\begin{aligned} \chi_{\text{D}} &= \frac{1-R'_{\text{d}}}{R'_{\text{d}}} a_{\text{R}} \\ &= \frac{1-(1-\gamma)R_{\text{d}}}{(1-\gamma)R_{\text{d}}} a_{\text{R}} \, . \end{aligned} \tag{70}$$

In the following, we derive a lower bound on the decoding complexity of a sequence of ensembles of RP-LDPC codes.

**Theorem 6.2.** Let $\{(n_r, \lambda, \rho)\}_{r=1}^{\infty}$ be a sequence of ensembles of LDPC codes whose block length $(n_r)$ tends to infinity as $r \to \infty$. Assume that a sequence of ensembles of RP-LDPC codes is constructed in the following way: for each code from an ensemble of the original sequence, a subset of $\alpha n_r$ code bits is a-priori selected, and these bits are randomly punctured at a fixed rate $(P_{\text{pct}})$. Assume that the punctured codes are transmitted over an MBIOS channel with capacity $C$, and that as $r$ tends to infinity, the sequence of ensembles of punctured codes achieves a fraction $1-\varepsilon$ of the capacity with vanishing bit error probability. Then in probability 1 w.r.t. the random puncturing patterns, the decoding complexity of this sequence under MPI decoding satisfies

$$\chi_{\text{D}}(\varepsilon) \geq K_1 + K_2 \ln\left(\frac{1}{\varepsilon}\right) \tag{71}$$

where

$$K_1 = -\frac{(1-\overline{C}) \, \ln\left(\frac{1}{2\ln 2}\frac{1-\overline{C}}{\overline{C}}\right)}{\overline{C} \, \ln\big((1-P_{\text{pct}}+\xi)g_1\big)}, \quad K_2 = -\frac{1-\overline{C}}{\overline{C} \, \ln\big((1-P_{\text{pct}}+\xi)g_1\big)} \tag{72}$$

where $g_1$ is introduced in (46), $\xi$ is introduced in (48), and $\overline{C} \triangleq (1-\alpha P_{\text{pct}})C$. For the particular case of a BEC, the term $\frac{1}{2\ln 2}$ can be dropped, thus improving the tightness of the additive term $(K_1)$ in the lower bound.

*Proof.* Since the code bits of a subset of the code bits whose size is $\alpha n_r$ are randomly punctured at rate $P_{\text{pct}}$, then the average puncturing rate is given by $\gamma = \alpha P_{\text{pct}}$. Hence, Eq. (68) yields that $\overline{C} = (1 - \alpha P_{\text{pct}})C$. By multiplying both sides of (47) by $1 - \alpha P_{\text{pct}}$ and getting from (69) that $R_{\text{d}} = (1 - \varepsilon)C$, we obtain

$$(1-\varepsilon)\overline{C} \leq 1 - \frac{1-\overline{C}}{1 - \frac{1}{2\ln 2}\sum_{p=1}^{\infty}\left\{\frac{1}{p(2p-1)}\,\Gamma\Big((1-P_{\text{pct}}+\xi)g_p\Big)\right\}}.$$

Following the same steps as in the proof of Theorem 6.1, we get a lower bound on the average right degree of the bipartite graph which corresponds to the pair of degree distributions $(\lambda, \rho)$. This lower bound is of the form

$$a_{\text{R}} > K_1' + K_2' \ln\left(\frac{1}{\varepsilon}\right) \tag{73}$$

where

$$K_1' = -\frac{\ln\left(\frac{1}{2\ln 2}\frac{1-\overline{C}}{\overline{C}}\right)}{\ln\Big((1-P_{\text{pct}}+\xi)g_1\Big)}, \quad K_2' = -\frac{1}{\ln\Big((1-P_{\text{pct}}+\xi)g_1\Big)}.$$

Note that $K_2$ is positive; this follows from (48), which yields that $\xi < (1-\alpha)P_{\text{pct}}$ (due to the fact that the integral of $\lambda$ over the interval [0, 1] is upper bounded by $\frac{1}{2}$). This assures that as the gap (in rate) to capacity vanishes, the lower bound on $a_{\text{R}}$ scales like the logarithm of the inverse of this gap.

From (69), we get $R_{\text{d}}' = (1-\varepsilon)\overline{C} < \overline{C}$, and therefore $\chi_{\text{D}} = \frac{1-R_{\text{d}}'}{R_{\text{d}}'}\, a_{\text{R}} > \frac{1-\overline{C}}{\overline{C}}\, a_{\text{R}}$. The proof of the lower bound on the decoding complexity is completed by multiplying both sides of (73) by $\frac{1-\overline{C}}{\overline{C}}$. In the particular case where the communication channel is a BEC, following the same concept as in the proof of Theorem 6.1 leads to the improved coefficient $K_1$. □

The upper bound on the decoding complexity for sequences of ensembles of IP-LDPC codes is also given in terms of the gap between the rate of the punctured rate and the channel capacity.

**Theorem 6.3.** Let $\{(n_r, \lambda, \rho, \pi^0)\}_{r=1}^{\infty}$ be a sequence of ensembles of IP-LDPC codes transmitted over an MBIOS channel whose capacity is $C$. If this sequence achieves a fraction $1-\varepsilon$ of the capacity with vanishing bit error probability, then in probability 1 w.r.t. the random puncturing patterns, the decoding complexity of this sequence under MPI decoding satisfies

$$\chi_{\text{D}}(\varepsilon) \geq K_1 + K_2 \ln\left(\frac{1}{\varepsilon}\right) \tag{74}$$

where

$$K_1 = -\frac{(1-\overline{C})\,\ln\left(\frac{1}{2\ln 2}\frac{1-\overline{C}}{\overline{C}}\right)}{\overline{C}\,\ln\Big(\left(1-\sum_{j=1}^{\infty}\lambda_j\pi_j\right)\,g_p\Big)}, \quad K_2 = -\frac{1-\overline{C}}{\overline{C}\,\ln\Big(\left(1-\sum_{j=1}^{\infty}\lambda_j\pi_j\right)\,g_p\Big)} \tag{75}$$

where $g_1$ is introduced in (46), and $\overline{C} \triangleq (1 - \sum_{j=1}^{\infty}\Lambda_j\pi_j)C$. For the particular case of a BEC, the term $\frac{1}{2\ln 2}$ can be dropped, thus improving the tightness of the additive term ($K_1$) in the lower bound.

*Proof.* The proof follows from the same concepts as the proof of Theorem 6.2, but is based on (56) instead of (47). Note that $K_2$, which reflects the logarithmic growth rate of the lower bound in (74), is always positive; this follows from (75) and due to the fact that from (46), $g_1 < 1$, and also $0 < 1 - \sum_{j=1}^{\infty}\lambda_j\pi_j \leq 1$. □

### 6.3 Re-Derivation of Reported Lower Bounds on the Decoding Complexity

In [10, Theorems 3 and 4], Pfister et al. introduced lower bounds on the decoding complexity of punctured codes on graphs with iterative decoding. The bounds were derived for the case where a subset of linearly independent code bits whose size is equal to the code dimension are randomly punctured at a fixed rate ($P_{\text{pct}}$), and the transmission of the codes takes place over an MBIOS channel. In particular, this scenario corresponds to RP-LDPC codes (see Section 5.2) where we choose a subset of the code bits to be randomly punctured at rate $P_{\text{pct}}$; under the assumption in [10, Theorems 3 and 4], the fraction ($\alpha$) of the code bits which are randomly punctured is equal to the code rate. In the appendix, we show that for randomly punctured LDPC codes, the lower bounds on the decoding complexity given in [10, Theorems 3 and 4] follow from a looser version of the bound in Theorem 6.2.

## 7 Summary and Outlook

The main result in this paper, Theorem 4.1, provides an upper bound on the asymptotic rate of a sequence of ensembles of LDPC codes which achieves vanishing bit error probability. We assume that the communication takes place over a set of parallel memoryless binary-input output-symmetric (MBIOS) channels. The derivation of Theorem 4.1 relies on upper and lower bounds on the conditional entropy of the transmitted codeword given the received sequence at the output of the parallel channels (see Section 3), and it is valid under optimal ML decoding (or any sub-optimal decoding algorithm). This theorem enables the derivation of a lower bound on the decoding complexity (per iteration) of ensembles of LDPC codes under message-passing iterative decoding when the transmission of the codes takes place over parallel MBIOS channels. The latter bound is given in terms of the gap between the rate of these codes for which reliable communication is achievable and the channel capacity. Similarly to a lower bound on the decoding complexity of ensembles of LDPC codes for a single MBIOS channel [15], the lower bound on the decoding complexity which is derived for parallel channels also grows like the log of the inverse of the gap to capacity.

Theorem 4.1 can be used for various applications which form particular cases of communication over parallel channels, e.g., intentionally punctured LDPC codes [4], non-uniformly error protected LDPC codes [11], and LDPC-coded modulation (see e.g., [6, 16]). In Section 5, we use Theorem 4.1 for the derivation of upper bounds on the achievable rates under ML decoding of (randomly and intentionally) punctured LDPC codes whose transmission takes place over an MBIOS channel. It is exemplified numerically that for various good ensembles of IP-LDPC codes, the asymptotic loss in performance due to the code structure is still non-negligible as compared to the corresponding loss due to the sub-optimality of iterative decoding (as compared to optimal ML decoding). Looser versions of the bounds derived in this paper for punctured LDPC codes suggest a simplified re-derivation of previously reported bounds on the decoding complexity of randomly punctured LDPC codes (see [10, Theorems 3 and 4]).

Interconnections between the theorems introduced in this paper and some other previously reported results are shown in Fig. 1 (see p. 32).

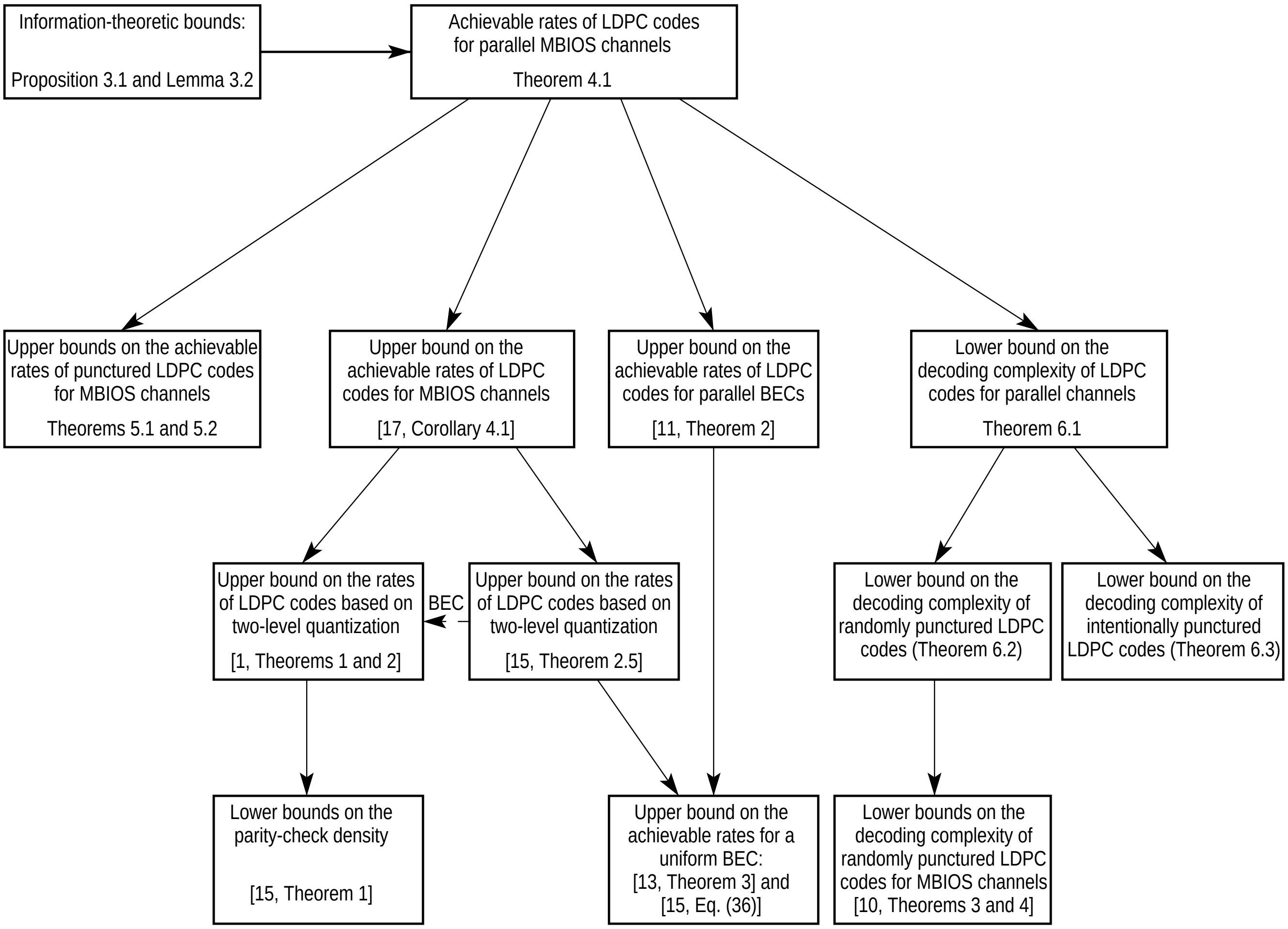

Figure 1: An interconnections diagram among the bounds in this paper and previously reported bounds.

### Acknowledgment

I. Sason wishes to acknowledge Henry Pfister and Ruediger Urbanke for stimulating discussions on puncturing theorems during the preparation of the work in [10]. The work of I. Sason was supported by the Taub and Shalom Foundations.

## Appendix: Re-derivation of [10, Theorems 3 and 4]

In the following, we start with the re-derivation of [10, Theorem 4] for general MBIOS channels, and then re-derive the refined bound in [10, Theorem 3] for a BEC. For the re-derivation of [10, Theorems 3 and 4] we rely on Theorem 6.2 whose derivation is based on Theorem 5.1. Hence, we first loosen the upper bound on the achievable rates given in (47), and then re-derive [10, Theorem 4] as a consequence of this looser version. The loosening of (47) is done by replacing the positive parameter $\xi$ introduced in (48) to zero, and then using the lower bound on $\Gamma$ from

Lemma 6.1. This gives

$$R_{\rm d} \le \frac{1}{1-\alpha P_{\rm pct}} \left( 1 - \frac{1-(1-\alpha P_{\rm pct})C}{1 - \frac{1}{2\ln 2} \sum_{p=1}^{\infty} \left\{ \frac{1}{p(2p-1)} \left( (1-P_{\rm pct}) g_p \right)^{a_{\rm R}} \right\}} \right) . \tag{A.1}$$

Finally, truncating the infinite series in the RHS of (A.1) by only taking its first term which corresponds to $p = 1$ further loosens the upper bound on the achievable rates, and gives

$$R_{\rm d} \le \frac{1}{1-\alpha P_{\rm pct}} \left( 1 - \frac{1-(1-\alpha P_{\rm pct})C}{1 - \frac{1}{2\ln 2}\left( (1-P_{\rm pct}) g_1 \right)^{a_{\rm R}}} \right) . \tag{A.2}$$

From (69), we get the inequality

$$(1-\varepsilon)(1-\alpha P_{\rm pct})C \le 1 - \frac{1-(1-\alpha P_{\rm pct})C}{1 - \frac{1}{2\ln 2}\left( (1-P_{\rm pct}) g_1 \right)^{a_{\rm R}}}.$$

which after straightforward algebra gives

$$1 + \frac{1-(1-\alpha P_{\rm pct})C}{\varepsilon C (1-\alpha P_{\rm pct})} \le 2\ln 2 \left( \frac{1}{(1-P_{\rm pct}) g_1} \right)^{a_{\rm R}} . \tag{A.3}$$

We proceed by giving a simple lower bound on $g_1$.

**Lemma A.1.** For $g_1$ introduced in (46), the following inequality holds

$$g_1 \ge (1-2w)^2$$

where

$$w \triangleq P_{\rm e}(a) = \frac{1}{2} \Pr(L=0) + \int_{-\infty}^{0^-} a(l)\, dl$$

designates the uncoded bit error probability of the MBIOS channel given the channel input is 1.

*Proof.* Based on the symmetry property where $a(l) = e^l a(-l)$ and Jensen's inequality, we get

$$\begin{aligned}
g_1 &= \int_0^{\infty} a(l)\, (1+e^{-l}) \tanh^2\left(\frac{l}{2}\right) dl \\
&= \int_{-\infty}^{\infty} a(l) \tanh^2\left(\frac{l}{2}\right) dl \\
&\ge \left( \int_{-\infty}^{\infty} a(l) \tanh\left(\frac{l}{2}\right) dl \right)^2 \\
&= \left( \int_0^{\infty} a(l)\, (1+e^{-l}) \tanh\left(\frac{l}{2}\right) dl \right)^2 \\
&= \left( \int_0^{\infty} a(l)\, (1-e^{-l}) dl \right)^2 \\
&= \left( \int_{0^+}^{\infty} \big(a(l) - a(-l)\big) dl \right)^2 \\
&= \left( 1 - \Pr(L=0) - 2\int_{-\infty}^{0^-} a(l) dl \right)^2 \\
&= (1-2w)^2.
\end{aligned}$$

□

Dropping the 1 in the LHS of (A.3) and replacing $g_1$ in the RHS of (A.3) by its lower bound from Lemma A.1 gives

$$\begin{aligned} \frac{1-(1-\alpha P_{\text{pct}})C}{\varepsilon C(1-\alpha P_{\text{pct}})} &\leq 2\ln 2\left(\frac{1}{(1-P_{\text{pct}})(1-2w)^2}\right)^{a_{\text{R}}} \\ &\leq 2\ln 2\left(\frac{1}{(1-P_{\text{pct}})(1-2w)}\right)^{2a_{\text{R}}}. \end{aligned}$$

Solving the last inequality for $a_{\text{R}}$ gives

$$a_{\text{R}} \geq \frac{\ln\left(\frac{1}{2\ln 2}\frac{1-(1-\alpha P_{\text{pct}})C}{\varepsilon C(1-\alpha P_{\text{pct}})}\right)}{2\ln\left(\frac{1}{(1-P_{\text{pct}})(1-2w)}\right)}.$$

Based on the equality (70) which relates the complexity under MPI decoding to the average right degree ($a_{\text{R}}$) and since $R_{\text{d}} < C$, we get from the last inequality

$$\chi_{\text{D}}(\varepsilon) \geq \frac{1-C}{2C}\,\frac{\ln\left(\frac{1}{2\ln 2}\frac{1-(1-\alpha P_{\text{pct}})C}{\varepsilon C(1-\alpha P_{\text{pct}})}\right)}{\ln\left(\frac{1}{(1-P_{\text{pct}})(1-2w)}\right)}\,. \tag{A.4}$$

Note that $\alpha = R_{\text{d}}$ in [10, Theorem 4]. This gives the equality $\alpha = (1-\varepsilon)(1-\alpha P_{\text{pct}})C$ whose solution is

$$\alpha = \frac{(1-\varepsilon)C}{1+(1-\varepsilon)CP_{\text{pct}}}\,. \tag{A.5}$$

Finally, the substitution of $\alpha$ in (A.5) into the RHS of (A.4) gives

$$\begin{aligned} \chi_{\text{D}}(\varepsilon) &\geq \frac{1-C}{2C}\,\frac{\ln\left(\frac{1}{\varepsilon}\frac{1-(1-P_{\text{pct}})C+\varepsilon CP_{\text{pct}}}{2C\ln 2}\right)}{\ln\left(\frac{1}{(1-P_{\text{pct}})(1-2w)}\right)} \\ &\geq \frac{1-C}{2C}\,\frac{\ln\left(\frac{1}{\varepsilon}\frac{1-(1-P_{\text{pct}})C}{2C\ln 2}\right)}{\ln\left(\frac{1}{(1-P_{\text{pct}})(1-2w)}\right)}\,. \end{aligned} \tag{A.6}$$

which coincides with [10, Theorem 4] for a sequence of ensembles of randomly punctured LDPC codes.

For the derivation of the refined bound for the BEC which is given in [10, Theorem 3], we start from (A.1). The refinement of the latter bound is due to the fact that for the BEC, $g_p$ in (46) is independent of $p$, and is equal to $g_p = 1 - P_{\text{BEC}}$ where $P_{\text{BEC}}$ designates the erasure probability of the BEC. From (A.1), we get the following upper bound on the achievable rates:

$$R_{\text{d}} \leq \frac{1}{1-\alpha P_{\text{pct}}}\left(1 - \frac{1-(1-\alpha P_{\text{pct}})C}{1-\Big((1-P_{\text{pct}})(1-P_{\text{BEC}})\Big)^{a_{\text{R}}}}\right)$$

which follows from the equality $\sum_{p=1}^{\infty}\frac{1}{2p(2p-1)} = \ln 2$. Substituting $R_{\text{d}} = (1-\varepsilon)(1-P_{\text{BEC}})$ and the $\alpha$ in (A.5) gives a lower bound on $a_{\text{R}}$. Finally, the lower bound in [10, Theorem 3] follows from the resulting lower bound on $a_{\text{R}}$ and the inequality $\chi_D(\varepsilon) \geq \frac{1-C}{C}\, a_{\text{R}}$.